\documentclass[]{mn2e}

\usepackage{amssymb}
\usepackage{amsmath}
\usepackage{graphicx}
\usepackage{color}
\usepackage{longtable}
\usepackage{url}

\title[Impact of M$_*$/L gradients on stellar masses]{M$_*$/L gradients driven by IMF variation: Large impact on dynamical stellar mass estimates}
\author[Bernardi et al.]{\parbox{\textwidth}{M. Bernardi$^{1}$\thanks{E-mail: bernardm@sas.upenn.edu}, R. K. Sheth$^{1}$, H. Dominguez-Sanchez$^{1}$, J.-L. Fischer$^{1}$, K.-H. Chae$^{1,2}$, M. Huertas-Company$^{1,3}$, F. Shankar$^{4}$} \vspace{0.4cm}\\
\parbox{\textwidth}{$^{1}$Department of Physics and Astronomy, University of Pennsylvania, Philadelphia, PA 19104, USA\\
$^{2}$ Department of Astronomy and Space Science, Sejong University, 98 Gunja-dong Gwangjin-Gu, Seoul 143-747, Republic of Korea\\ 
$^{3}$GEPI, Observatoire de Paris, CNRS, Univ. Paris Diderot;
Place Jules Janssen, 92190 Meudon, France\\
$^{4}$Department of Physics and Astronomy, University of Southampton,
Southampton SO17 1BJ, UK\\}}

\begin{document}
 \date{Accepted .  Received ; in original form }

\maketitle

\label{firstpage}

\begin{abstract}
  Within a galaxy the stellar mass-to-light ratio $\Upsilon_*$ is not constant.
 Recent studies of spatially resolved kinematics of nearby early-type galaxies suggest that allowing for a variable initial mass function (IMF) returns significantly larger $\Upsilon_*$ gradients than if the IMF is held fixed. We show that ignoring such IMF-driven $\Upsilon_*$ gradients can have dramatic effect on dynamical ($M_*^{\rm dyn}$), though stellar population ($M_*^{\rm SP}$) based estimates of early-type galaxy stellar masses are also affected.  This is because $M_*^{\rm dyn}$ is usually calibrated using the velocity dispersion measured in the central regions (e.g. $R_e/8$) where stars are expected to dominate the mass (i.e. the dark matter fraction is small). On the other hand, $M_*^{\rm SP}$ is often computed from larger apertures (e.g. using a mean $\Upsilon_*$ estimated from colors). If $\Upsilon_*$ is greater in the central regions, then ignoring the gradient can overestimate $M_*^{\rm dyn}$ by as much as a factor of two for the most massive galaxies. Large $\Upsilon_*$-gradients have four main consequences:
  First, $M_*^{\rm dyn}$ cannot be estimated independently of stellar population synthesis models.
 Second, if there is a lower limit to $\Upsilon_*$ and gradients are unknown, then requiring $M_*^{\rm dyn}=M_*^{\rm SP}$ constrains them.
    Third, if gradients are stronger in more massive galaxies, then accounting for this reduces the slope of the correlation between $M_*^{\rm dyn}/M_*^{\rm SP}$ of a galaxy with its velocity dispersion. In particular, IMF-driven gradients bring $M_*^{\rm dyn}$ and $M_*^{\rm SP}$ into agreement, not by shifting $M_*^{\rm SP}$ upwards by invoking constant bottom-heavy IMFs, as advocated by a number of recent studies, but by revising $M_*^{\rm dyn}$ estimates in the literature downwards.
    Fourth, accounting for $\Upsilon_*$ gradients changes the high-mass slope of the stellar mass function $\phi(M_*^{\rm dyn})$, and reduces the associated stellar mass density. These conclusions potentially impact estimates of the need for feedback and adiabatic contraction, so our results highlight the importance of measuring $\Upsilon_*$ gradients in larger samples.  
\end{abstract}


\section{Introduction}
A census of the stellar masses of galaxies is more useful than a census of their luminosity (e.g. Cole et al. 2001).  There are two approaches to estimating the stellar mass of a galaxy.  One fits the observed data -- single or multi-band photometry and perhaps spectroscopy as well -- to stellar population synthesis libraries (SP; e.g. Bruzual \& Charlot 2003; Maraston 2005; Vazdekis et al. 2010; Conroy \& Gunn 2010; Maraston \& Str{\"o}mb{\"a}ck 2011).  These include assumptions about the age, metallicity, star formation history, the amount of dust, of gas, and the initial mass function (IMF) (e.g. Worthey 1994; Thomas et al. 2011; Mendel et al. 2014; Villaume et al. 2017).  The data constrain (some, usually degenerate, combination of) these parameters, and hence the stellar mass-to-light ratio, $M_*^{\rm SP}/L$.  The stellar mass $M_*^{\rm SP}$ is then obtained by multiplying $M_*^{\rm SP}/L$ by an estimate of the total light.  Note that here it is the total $L$ which matters; the detailed shape of the light profile does not.  Both $M_*^{\rm SP}/L$ and $L$ carry significant uncertainties, although recent work suggests that the systematics associated with $L$ are now subdominant (Bernardi et al. 2017a).  The biggest systematic unknown for $M_*^{\rm SP}/L$ is the IMF.

The other is a dynamical estimate, $M_*^{\rm dyn}$, and comes in two flavors, both of which use the Jeans equation (see Cappellari 2016 for a review).  These estimates almost always assume that $M_*^{\rm dyn}/L$ is constant throughout the galaxy, so that the observed shape of the light profile is used as an indicator of the mass profile.  If the velocity dispersion is only measured on a single scale (e.g. the SDSS), then $M_*^{\rm dyn}\propto R_e\sigma^2/G$, where the size $R_e$ and the constant of proportionality depend on the shape of the light profile -- but the total $L$ does not matter.  In this case, the Jeans equation predicts the shape of the velocity dispersion profile, and the overall amplitude is got by matching the predicted shape to the observed $\sigma$ (this is why $L$ does not matter).  If spatially resolved spectroscopy is available, so the profile $\sigma(r)$, rather than its value at only a single scale, is known, then the richer dataset allows one to constrain a richer family of models.  Now the models must match $\sigma$ over a range of scales.  Typically, $\sigma(r)$ predicted by the light (Binney \& Mamon 1982; Prugniel \& Simien 1997) falls more steeply than observed (J{\o}rgensen et al. 1995), so the difference is attributed to dark matter which increasingly dominates the mass in the outer regions.  Importantly, in both cases, $M_*^{\rm dyn}$ does not depend on the details of the stellar population, so it is sometimes viewed as being less impacted by systematics than $M_*^{\rm SP}$.  

It has been known for some time that, if one assumes that all galaxies have the same IMF, then $M_*^{\rm dyn}/M_*^{\rm SP}$ varies across the early-type population (Bender, Burstein \& Faber 1992; Bernardi et al. 2003; Shankar \& Bernardi 2009).  Recent work has focussed on the fact that this ratio tends to increase with velocity dispersion (Cappellari et al. 2013 from ATLAS$^{\rm 3D}$; Li et al. 2017 from MaNGA).  This correlation is substantially reduced if the IMF is treated as a free parameter:  galaxies with larger $\sigma$ tend to have IMFs which result in larger $M_*^{\rm SP}/L$ (Conroy \& van Dokkum 2012; Lyubenova et al. 2016; Tang \& Worthey 2017).  In this case, the difference between $M_*^{\rm dyn}$ and $M_*^{\rm SP}$ is explained by asserting that $M_*^{\rm SP}/L$ estimates are biased unless one accounts for IMF-related effects (Cappellari et al. 2013; Li et al. 2017; but see Clauwens et al. 2016).  

While this is possible, it is potentially problematic for the following reason.  Recall that $M_*^{\rm SP} = (M_*^{\rm SP}/L)\times L$.  The improvements in $L$ are most dramatic for the most massive galaxies (Bernardi et al. 2010; Bernardi et al. 2013; Meert et al. 2015), and have lead to a revision of the stellar mass density upwards by a factor of 2 (Bernardi et al. 2013, 2017a,b; Thanjavur et al. 2015; D'Souza et al. 2015).  This weakens the role that must have been played by feedback in regulating star formation.  If the $M_*^{\rm SP}$ estimates must be increased further because of IMF effects on $M_*^{\rm SP}/L$ (Bernardi et al. 2018), then the need for feedback will be further reduced.  It is not obvious that this is reasonable.

We turn, therefore, to the possibility that some of the discrepancy between $M_*^{\rm dyn}$ and $M_*^{\rm SP}$ is driven by problems with $M_*^{\rm dyn}$.  For this study, we will suppose that the problems are not due to shape of the light profile, but to the assumption that the stellar mass-to-light ratio $\Upsilon_*$ is constant within a galaxy.  The constant $\Upsilon_*$ assumption is commonly made (e.g. the analysis of ATLAS$^{\rm 3D}$ in Cappellari et al. 2013) because it is convenient -- it has no physical motivation.  Galaxies have long been known to show gradients in color (photometry) and absorption line strength (spectroscopy).  In the context of stellar population modelling, these indicate age and/or metallicity gradients.  Typically, these gradients tend to increase the stellar population estimate of the stellar mass-to-light ratio in the central regions. 
Age and metallicity estimates which result from assuming a constant IMF imply a variation in $\Upsilon_*$ of about 50\% within a galaxy (e.g. bottom row, second from left panel in Figure~10 of van Dokkum et al. 2017; also see Newman et al. 2015, who report $\Upsilon_*\propto R^{-0.15}$ in massive galaxies).  When luminosity-weighted and averaged over the galaxy this is a smaller effect, which is why ignoring these gradients when estimating $M_*^{\rm dyn}$ may not be so problematic. Indeed, in their study of $M_*^{\rm dyn}$ in MaNGA, Li et al. (2017) report that the impact of such fixed-IMF $\Upsilon_*$-gradients on $M_{\rm dyn}$ estimates is small (see their Figure~6).

However, more recent work suggests that IMF-sensitive line-indices also show gradients, which indicate that the IMF is more bottom-heavy in the center than it is beyond the half-light radius (Mart{\'i}n-Navarro et al. 2015; Lyubenova et al. 2016; van Dokkum et al. 2017; La Barbera et al. 2017; Li et al. 2017; Parikh et al. 2018), although there is not yet universal agreement (Alton et al. 2017; Vaughan et al. 2017).  These tend to further increase the stellar population-based $M_*^{\rm SP}/L$ estimate in the central regions.  The panel which is second from bottom right of Figure~10 of van Dokkum et al. (2017) suggests that the net effect can be a factor of 3 or more, meaning that IMF-gradients are the dominant contribution to gradients in $\Upsilon_*$.  Importantly, a factor of 3 variation is too large to be safely ignored when using the Jeans-equation.  

van Dokkum et al.'s conclusions are based on only 6 objects.  However, Parikh et al. (2018) study a much larger sample (by more than two orders of magnitude), drawn from the MaNGA survey, and they too see IMF-related gradients.  The IMF gradients they report are slightly weaker than those in van Dokkum et al., and substantially weaker at smaller masses, but they are large enough that they should not be ignored.  

In Section~2 we present the first analysis of how such IMF-driven $M_*/L$ gradients impact Jeans-equation estimates of $M_*^{\rm dyn}$.  In Section~3 we show how recalibrated masses modify the $M_*^{\rm dyn}/M_*^{\rm SP}$ scaling, as well as the stellar mass function.  A final section summarizes and discuss consequences for estimates of dark matter fractions and evidence for/against adiabatic contraction.  An Apppendix provides a fully analytic toy model which illustrates the main features of our results.

When necessary, we assume a spatially flat background cosmology with parameters $(\Omega_m,\Omega_\Lambda)=(0.3,0.7)$, and a Hubble constant at the present time of $H_0=70$~km~s$^{-1}$Mpc$^{-1}$, as these are the values adopted in most studies of the stellar mass function which we reference in our work.  As we will be working at low $z$, all our conclusions are robust to small changes in these parameters.

\section{Effect of $\Upsilon_*$ gradients on stellar mass estimates}
We study the effect of gradients in the observed (2d, projected) stellar mass-to-light ratio, $\Upsilon_*$.  We show that increasing $\Upsilon_*$ in the central regions has a much more dramatic effect on $M_*^{\rm dyn}$ than on $M_*^{\rm SP}$.  Ignoring the gradient leads to an over-estimate of $M_*^{\rm dyn}$.  Much of the analysis here is numerical.  See Appendix~A for a toy model which makes use of simple analytic approximations to the three-dimensional profiles of galaxies to illustrate these same points.

\subsection{Observationally motivated scalings}
We begin with the projected light distribution, $I(R)$, which is observed to follow a S{\'e}rsic (1963) profile:  
\begin{equation}
  I(R) = \frac{L}{R_e^2}\,\frac{b^{2n}}{2\pi\, n\Gamma(2n)}\,
         {\rm e}^{-b_n\,(R/R_e)^{1/n}},
\end{equation}
with $b_n\approx 2n - 1/3 + 0.01/n$.

For the projected mass-to-light ratio we set
\begin{equation}
 \Upsilon_*(R) = \Upsilon_{*0}\, (1+\alpha - \beta\, R/R_e)  \quad {\rm if}\quad R/R_e\le \alpha/\beta,
 \label{Ur}
\end{equation}
with $\Upsilon_*(R) = \Upsilon_{*0}$ at larger $R$.  The values $(\Upsilon_{*0},\alpha,\beta) = (3,2.33,6)$ provide a good approximation to the scalings shown in the panel which is second from right in the bottom of Figure~10 of van Dokkum et al. (2017).

The total mass associated with this gradient is
\begin{equation}
  M_\infty \equiv 2\pi \int_0^\infty {\rm d}R\,R\,J(R)
              = \Upsilon_{*0} L\,\Bigl(1 + g(n,\alpha,\beta)\Bigr),
 \label{Minf}
\end{equation}
where
\begin{equation}
 J(R) \equiv I(R)\,\Upsilon_*(R).
 \label{JR}
\end{equation}
It is useful to think of the second term, 
\begin{equation}
  M_{\rm grad}
  \equiv \Upsilon_{*0} L \, g(n,\alpha,\beta),
 \label{Madded}
\end{equation}
as the extra mass contributed by the gradient.  
For $\Upsilon_*(R)$ given by equation~(\ref{Ur}), $M_{\rm grad}$ can be written in terms of incomplete gamma functions.  For $n=6$ (Meert et al. 2015 show that this is typical for the massive galaxies, $M_*^{\rm SP} > 10^{11}$ M$_{\odot}$, of most interest in this paper), and the values of $\alpha$ and $\beta$ given above, $M_{\rm grad} = 0.4\,\Upsilon_{*0} L$; the gradient term contributes an additional 40\% to the total mass.  The ratio $M_{\rm grad}/\Upsilon_{*0} L$ decreases for smaller $n$.

In what follows, we will illustrate our results using three choices for the gradient strength.  These are summarized in Table~\ref{tab:ab}.  The smallest values are chosen to mimic the gradient associated with fixed-IMF (panel which is second from bottom left in Figure~10 of van Dokkum et al. 2017), and the largest are the values we just described (second from bottom right of Figure~10 in van Dokkum et al. 2017).  We also study an intermediate case, in which the IMF in the central regions cannot become arbitrarily bottom heavy, but is capped at Salpeter (1995):  one may think of this as a simple approximation to the gradients reported by Parikh et al. (2018), but we caution that they provide a number of different estimates of IMF-gradients which vary widely.  While our three models have different slopes for $\Upsilon_*$, the transition scale beyond which $\Upsilon_*$ becomes constant (i.e. $\Upsilon_{*0}$) is the same:  it equals $\alpha/\beta\sim 0.4 R_e$. We have not explored models where this scale is also varied, simply because varying the slope alone allows us to illustrate all the important steps in the argument.

\subsection{Stellar population estimate}
The presence of gradients in $\Upsilon$ complicate interpretation of published $M_*^{\rm SP}$ values.  To illustrate why, we consider various ways in which $M_*^{\rm SP}$ may have been estimated.  These depend on the answer to two questions:  Was the IMF treated as a free parameter? Was spatially resolved information used?

For the most common estimates, the answer to both questions is `no'.  For such fixed-IMF estimates, typically based on integrated multi-band photometry, what is returned is $\Upsilon_{*\rm fix}$, which is then multiplied by the luminosity to give $M_{\infty,{\rm fix}}$.  In the current context, this would correspond to $M_\infty$ of equation~(\ref{Minf}) with $\alpha,\beta$ values from the first row in Table~\ref{tab:ab} (i.e. Model $=$ fixed IMF).  
The corresponding estimate when the IMF is allowed to vary is given by either of the other two pairs in the Table.  In general, these will be larger than the one for a fixed IMF.  

Often, however, $\Upsilon_*$ is estimated from spectroscopic measurements (line indices, etc.) which probe scales that are of order $R_e$ or smaller.  A particularly relevant example in the current context is Conroy \& van Dokkum (2012); their IMF estimates were based on the light within $R_e/8$.  If we use $R_{\rm obs}$ to denote this scale,then the reported $M_*^{\rm SP}$ is given by 
\begin{equation}
 M_*^{\rm SP}\approx \frac{2\pi \int_0^{R_{\rm obs}} {\rm d}R'\,R'\,J(R')}{2\pi \int_0^{R_{\rm obs}} {\rm d}R'\,R'\,I(R')}\,\times\, L ,
\end{equation}
and will be greater than $M_\infty$.  Hence, if the IMF was a free parameter, then the estimate from within $R_{\rm obs}$ is an overestimate.  On the other hand, if the IMF was held fixed, then the overestimate relative to $M_{\infty,{\rm fix}}$ might still be smaller than the true $M_\infty$ (the one when IMF-gradients are allowed), because $M_{\infty,{\rm fix}}<M_\infty$.  We show examples of this in Section~\ref{impact}.

\begin{table}
  \centering
  \caption{Parameter values for M/L gradient strength (equation~\ref{Ur}) driven by different assumptions about the IMF (fixed or variable) used in this study.}
  \begin{tabular}{lcccl}
  \hline
  Model         & IMF-var. &    $\alpha$      & $\beta$ & Source\\
  \hline
  fixed IMF     &    no    &             0.39 &   1.00  & Chabrier\\
  Salp$^{\rm IN}$-Chab$^{\rm OUT}$ & yes  &  1.29 &   3.33  & Salpeter-Chabrier \\
  vD17                         &  yes &  2.33 &   6.00 & van Dokkum et al.\\
  \hline
  \end{tabular}
  \label{tab:ab}
\end{table}

\subsection{Dynamical estimate}
Dynamical mass estimates are also affected.  For $M_*^{\rm dyn}$, the most relevant quantities are the magnitude of $M_{\rm grad}$ (equation~\ref{Madded}) and its spatial distribution.  Recall that, for the numbers given earlier (the model with the strongest gradient, i.e. Model $=$ vD17 in Table~\ref{tab:ab}), $M_{\rm grad}$ contributes an additional 40\% to the total mass.  However, this additional mass is entirely within $0.4R_e$.  Absent gradients, the mass with $0.4R_e$ equals about $0.3\,\Upsilon_{*0} L$, so the gradient more than doubles the mass within $0.4R_e$.  Since the dynamical mass is estimated by normalizing to $\sigma$ measured within $R_e/8$ or so, the impact of this additional mass on the $M_{\rm dyn}$ estimate will be much greater than 40\%, and more like a factor of $2$, as we now quantify.

Following, e.g., Binney \& Mamon (1982), the three dimensional stellar mass profile is obtained by deprojecting $J(R)\equiv \Upsilon_*(R)\,I(R)$, the projected stellar mass profile:  
\begin{equation}
  \rho_{\rm *}(r)
 = -\frac{1}{\pi}\int_r^\infty {\rm d}R\,\frac{{\rm d}J/{\rm d}R}{\sqrt{R^2-r^2}}.
\end{equation}
With $\rho_{\rm *}(r)$ in hand it is straightforward to obtain $M_*(<r)$, from which (integrating) the Jeans equation,
\begin{equation}
 \frac{{\rm d}\, \rho_{\rm *}(r)\sigma^2(r)}{{\rm d}r} + \frac{2\beta_\sigma(r)}{r}\, \rho_{\rm *}(r)\sigma^2(r) = -\rho_{\rm *}(r)\frac{GM(<r)}{r^2},
\end{equation}
yields $\sigma^2(r)$.  The quantity $\beta_\sigma(r)$ (not to be confused with our parametrization of the $\Upsilon$ gradient) is the velocity anisotropy parameter:  we set it to zero in what follows, and comment on this approximation at the end of this subsection.  The enclosed mass $M(<r)$ is the sum of that in stars $M_*(<r)$ and in dark matter $M_{\rm DM}(<r)$.  Although $M_{\rm DM}(<r)$ is not known, we expect $M(<r)\approx M_*(<r)$ on small enough scales.  Light-weighting (rather than mass-weighting) and projecting $\sigma^2(r)$ gives $\sigma^2_p(R)$:
\begin{equation}
  I(R)\,\sigma_p^2(R) = 2\int_R^\infty \frac{{\rm d}r\,r\,\rho_{\rm L}(r)\,\sigma^2(r)}{(r^2-R^2)^{1/2}} \left[1 - \beta_\sigma(r)\frac{R^2}{r^2}\right],
 \label{sigmap}
\end{equation}
where
\begin{equation}
  \rho_{\rm L}(r)
 = -\frac{1}{\pi}\int_r^\infty {\rm d}R\,\frac{{\rm d}I/{\rm d}R}{\sqrt{R^2-r^2}}.
\end{equation}
is the deprojected light profile.
Often, it is the velocity dispersion within an aperture which is observed.  This is 
\begin{equation}
 I(<R)\,\sigma^2_p(<R) \equiv 2\pi \int_0^R {\rm d}R\,R\,I(R)\,\sigma^2_p(R).
\end{equation}
The Jeans equation analysis treats $\Upsilon_{*0}$ -- the constant amplitude factor which appears as the first term on the right hand side of equation~(\ref{Ur}) -- as a free parameter which is fixed by matching the predicted $\sigma_p(<R)$ to that observed.  Note that, unless a model for the dark matter is explicitly included, this matching must be done on small enough scales that neglecting dark matter is accurate.  Once this has been done, then using $M_*(<r)$ in the Jeans equation even on larger scales yields an estimate of the contribution to the observed $\sigma_p(<r)$ which is due to the stars.

Although the overall normalization $\Upsilon_{*0}$ is a free parameter, the $R$-dependence of $\Upsilon_*(R)$, the steepness of the gradient, matters in what follows.  This steepness will obviously affect the scale dependence of $\sigma_p(R)$.  Therefore, anything else that changes the steepness of the predicted $\sigma_p(R)$ will also affect our results.  For example, to illustrate our arguments, we set $\beta_\sigma=0$, meaning we assume orbits are isotropic.  Figure~7 in Ciotti \& Lanzoni (1997) suggests that anistropies contribute less than 10\% effects for the E+S0s of most interest here. We explore this briefly in the next sub-Section, but a detailed study of anisotropy is beyond the scope of this work.  

\begin{figure}
 \centering
 \includegraphics[width=1.\hsize]{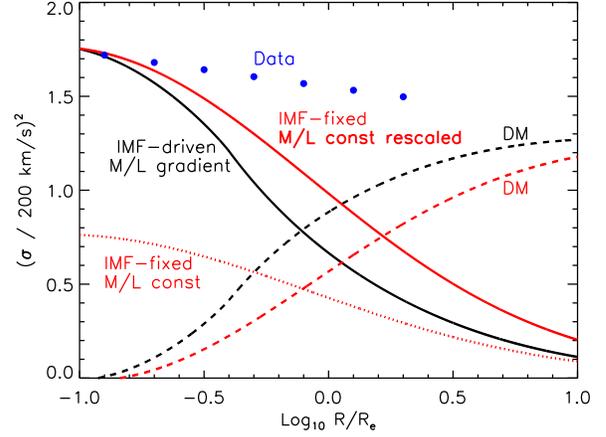}
 \caption{Effect of gradients in $\Upsilon_* = M_*^{\rm SP}/L$ (driven by IMF-gradients) on $M_*^{\rm SP}$ and $M_*^{\rm dyn}$ estimates.  Filled blue circles represent the measured velocity dispersion (equation~\ref{sigmacorr}).  Lower dotted red curve shows the predicted shape for a fixed IMF (e.g. Chabrier 2003) and a constant mass-to-light ratio when the light profile is S{\'e}rsic with $n=6$; upper solid red curve shows the $M_*^{\rm SP}/L$ that is required for this shape to fit the observed $\sigma^2$ on small scales.  The discrepancy between the solid red curve and the blue circles on larger scales (dashed red curve) is attributed to dark matter.  Black solid and dashed curves show the corresponding results if $M_*^{\rm SP}/L$ increases towards the center as given by equation~(\ref{Ur}) with $(\alpha,\beta)=(2.33,6)$.  As the stellar mass is now more centrally concentrated, the associated velocity dispersion falls more steeply from the center, so that more dark matter is needed within $R_e$ to explain why the observed $\sigma^2$ is relatively flat.
 }
 \label{grad2d}
\end{figure}

\subsubsection{Effect of gradients}\label{grad}
We begin with a cartoon of the effect.  The blue dots in Figure~\ref{grad2d} represent measurements in data:  they show the mean light-weighted projected velocity dispersion within an aperture of projected size $R_e$:
\begin{equation}
 \sigma_p(<R) = \sigma_p(<R_e)\, (R_e/R)^{0.06}
 \label{sigmacorr}
\end{equation}
(J{\o}rgensen et al. 1995; Cappellari et al. 2006).  The solid black curve shows $\sigma_p(<R)$ returned from our Jeans analysis.  We scaled its height so that it matches the blue dots at $R=R_e/8$, as this is a common choice.  (Normalizing on small $R$ is necessary to be as immune to dark matter as possible.)  The dynamical mass estimate follows from the procedure we described earlier.  The black curve falls more steeply than the dots, showing that the mass in stars cannot account for the velocity dispersion observed on larger scales.  The dashed curve shows the difference between the observed dispersion and that predicted by the stellar mass.  This difference is usually attributed to dark matter.  This additional contribution exceeds that from the stars beyond about $0.8R_e$. 

The red curve shows the corresponding estimate if we ignore gradients by setting $J(R)\propto I(R)$ and then follow all the same steps as before.  When gradients are included, then $\rho_{\rm *}(r)$ is steeper, and $\sigma_p(<R)$ is too.  Therefore, a smaller total dynamical stellar mass can account for the small scale $\sigma_p$ (because the mass is more centrally concentrated).

A crude estimate of the difference in the total stellar dynamical mass estimates  (i.e., $M_{*\rm dyn}$, not the total dynamical stellar+dark matter mass) can be got as follows.  Since $M_{*\rm dyn}(<R) \sim R \sigma^2(<R)/G$, the ratio of the red and black solid curves is approximately equal to the ratio of the two stellar mass estimates enclosed within $R$.  At $R\ll R_e$, the two are similar, since both models are normalized to match the small scale velocity dispersion (the blue dots at $R\ll R_e$).  However, the total stellar dynamical mass is $M_{*\rm dyn}(<R)$ as $R \to \infty$.  So we must compare the red and black curves, not at $R = R_e/8$ where they are equal (by design), but at $R\gg R_e$. (At these large $R$, both curves lie well-below the blue dots because dark matter matters and these solid curves only show the stellar component.) There, the two $M_{*\rm dyn}(<R)$ estimates are quite different:  the red is about $2\times$ larger than the black.  This indicates that incorrectly ignoring gradients overestimates the total (i.e., the $R\to\infty$) $M_{*\rm dyn}$ estimate by a factor of about 2.

Moreover, since the solid red curve falls less steeply than the black one, it is closer to the observations over a wider range of scales.  Therefore, the associated dark matter estimate (dashed red) is smaller: it only dominates the estimated contribution from the stars beyond about $1.8R_e$.  Thus, ignoring gradients will lead one to systematically underestimate the actual dark matter contribution on small scales.  The Appendix shows that these are generic trends.  

\begin{figure}
 \centering
 \includegraphics[width=1.\hsize]{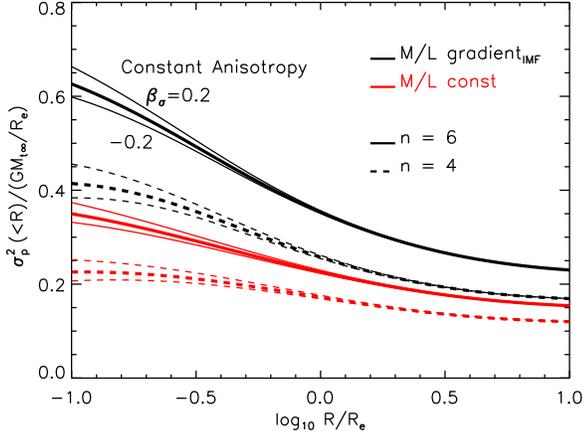}
 \caption{Effect of $\Upsilon_*$ gradients on the shape of the light-weighted velocity dispersion within a circular aperture of radius $R$, for three choices of the velocity anisotropy parameter $\beta_\sigma = (-0.2,0,0.2)$ when $n = 4$ (lower) and 6 (upper).  Red curves assume that mass is proportional to light, and black curves include a gradient in $\Upsilon_*$ (equation~\ref{Ur} with $\alpha=1.29$ and $\beta = 3.33$, i.e. Model $=$ Salp$^{\rm IN}$-Chab$^{\rm OUT}$ in Table~\ref{tab:ab}).  The three sets of curves for each $n$ show how a spatially constant velocity anisotropy $\beta_\sigma = (-0.2,0,0.2)$ affects the shape of the profile.  }
 \label{gradPS}
\end{figure}

Figure~\ref{gradPS} shows how these trends depend on the S{\'e}rsic index $n$ and on the whether or not the velocity dispersions are isotropic.  In anticipation of our consideration of the data in the next section, we first truncate the projected S{\'e}rsic profile at $7.5R_e$.
This reduces the total luminosity $L$ by an $n$-dependent amount, which is of order 10\%, and we use $M_{t\infty}$ to denote the total mass associated with this truncated profile. The figure shows two sets of black curves, and two sets of red:  solid and dashed curves show results for $n=4$ and 6.  In all cases the curves show $\sigma_p^2(<R)/(GM_{t\infty}/R_e)$, where $R_e$ is the projected half-light radius of the original (untruncated) profile.  The lower (red) curves show the profiles if there is no $\Upsilon_*$ gradient:  $(\alpha,\beta)=0$.  The upper (black) curves include a gradient, for which we used the intermediate set of values from Table~\ref{tab:ab}:  $(\alpha,\beta) = (1.29,3.33)$.  The central curve of each set is thicker; it assumes the velocity dispersion is isotropic.  The other two are for a constant anistropy of $\beta_\sigma=\pm 0.2$ at all $r$.  (As an aside, comparison of our red $\beta_\sigma=0$ curves with Figure~11 of Prugniel \& Simien 1997 shows that truncating the light profile makes only a small difference.)  While $\beta_\sigma\ne 0$ makes a visible difference, it is small compared to the effect of the $\Upsilon_*$ gradient.  Moreover, as we argue below, the real quantity of interest is the ratio of each black curve to its corresponding red one:  this removes most of the effect of $\beta_\sigma\ne 0$.

The estimated dynamical mass is 
\begin{equation}
  M_*^{\rm dyn} = k(R,n,\alpha,\beta)\, R_e\,\sigma_p^2(<\!R)/G, 
  \label{Mideal}
\end{equation}
where $k(R,n,\alpha,\beta)$ is the inverse of the predicted $\sigma^2_p(<\!R)/(GM_{t\infty}/R_e)$ on scale $R$ (e.g. Bernardi et al. 2018). Hence, the ratio of the black and red curves gives the amount by which one overestimates the total $M_*^{\rm dyn}$ if one ignores the gradient and normalizes to $\sigma_p(<\!R)$.  Note that $R$ appears in $k$ to highlight the fact that $k$ depends on the scale on which one chooses to normalize the model to the observed $\sigma_p(<\!R)$.  (If there were no dark matter, and the stellar mass profile and velocity anistropies were correctly modelled, then this dependence would make the estimated $M_*^{\rm dyn}$ the same for all choices of $R$.)  

\begin{figure}
 \centering
 \includegraphics[width=1.\hsize]{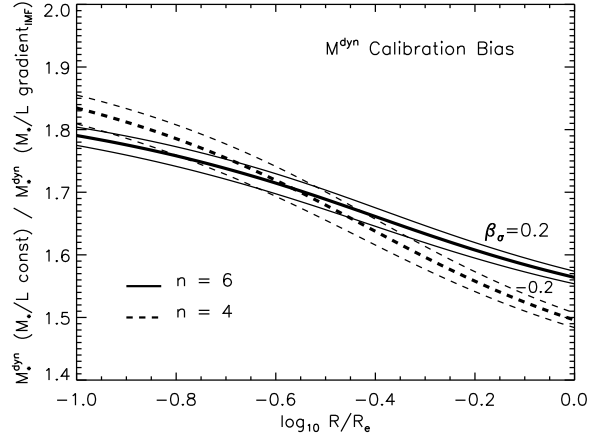}
 \caption{Bias in $M_{*\rm dyn}$ which results from ignoring $\Upsilon_*$ gradients, shown as a function of the scale on which the models are normalized to match the observed $\sigma^2_p(<\!R)$.  This bias is given by the ratio of the black to corresponding red curves in the previous figure.  When calibrating to the velocity dispersion measured inside $R_e/8$ -- where dark matter is not expected to contribute significantly -- the overestimate in $M^{\rm dyn}$ is $\sim 1.8$ for the Salp$^{\rm IN}$-Chab$^{\rm OUT}$ model (it is $\sim 2$ for the vD17 model). We do not show the ratio for $R>R_e$, because ignoring the effects of dark matter when calibrating to large scales is incorrect.}
 \label{biasMdyn}
\end{figure}

Figure~\ref{biasMdyn} shows this ratio for the various pairs of curves in Figure~\ref{gradPS}.  If $R\sim R_e/8$, the ratio is nearly a factor of $\sim 2$.  If, instead, one normalized at $R\sim R_e$, then this ratio is slightly smaller.  Recall, however, this ratio is only meaningful if dark matter makes a negligible contribution to $\sigma_p(<\!R)$.  If there are no gradients, then dark matter already contributes significantly at $R\sim R_e$; if gradients matter, then the scale where dark matter can be ignored is even smaller (c.f. Figure~\ref{grad2d}).  Therefore, only the $R\ll R_e$ values shown in Figure~\ref{biasMdyn} are likely to be reliable.

The steepening of the stellar contribution to $\sigma_p$ (because of $\Upsilon_*$ gradients) is what leads to a greater need for dark matter on small scales.  However, anistropic velocity dispersions can also affect the shape of $\sigma_p$. So, if $\beta_\sigma$ is not constrained by other observations, but is determined from the same Jeans equation analysis which determined $\Upsilon_{*0}$ and hence $M_*^{\rm dyn}$, then there is some room for degeneracy between anisotropy and dark matter fraction on small scales.  Previous work suggests this is a small effect (Ciotti \& Lanzoni 1997). The similarity of the different curves for each $n$ in Figure~\ref{biasMdyn} suggests that the effect of including anisotropic velocities is indeed small compared to the factor of two coming from the IMF-driven $\Upsilon_*$-gradient.  I.e., anisotropic velocity dispersions, if present, will not change our main point that current estimates of the dark matter fraction must be revisited if gradients matter.

To summarize: observations suggest that the stellar mass-to-light ratio $\Upsilon_*$ increases towards the center of a galaxy (equation~\ref{Ur} and Table~\ref{tab:ab}).  If these gradients are ignored, then
(i) stellar population-based mass estimates may be biased, and
(ii) dynamical mass estimates will be more strongly biased towards overestimating the total stellar mass and underestimating the amount of dark matter on small scales as a result.  In the next section, we use these results to illustrate the impact of gradients on the inferred stellar mass function.

\section{Potential impact of $\Upsilon_*$ gradients in SDSS}\label{impact}
This section illustrates the potential impact of $\Upsilon_*$ gradients using the galaxies in the SDSS DR7 Main Galaxy sample (Abazajian et al. 2009).

\subsection{The sample}\label{sdss}
We select the galaxies in the SDSS DR7 with r-band Petrosian magnitude limits $14 \le m_r \le 17.77$ mag (see Meert et al. 2015 for a detailed discussion of the sample selection), and we use the {\tt PyMorph} photometry of Meert et al. (2015).  The differences between {\tt PyMorph} and SDSS pipeline photometry are significant for the most massive galaxies.  See Bernardi et al. (2013), Meert et al. (2015), Fischer et al. (2017), and Bernardi et al. (2017b) for why {\tt PyMorph} is preferred.

For single S{\'e}rsic fits, the relevant {\tt PyMorph} parameters are the S{\'e}rsic index $n$, half-light radius $R_e$ and total luminosity $L$ of each object.  The estimated total light $L$ results from extrapolating the fitted (S{\'e}rsic) model to infinity.  As a result, the single S{\'e}rsic fits are known to slightly over-estimate the total light; integrating out to only $7.5R_e$ yields a more reliable luminosity estimate (Bernardi et al. 2017a; Fischer et al. 2017; Bernardi et al. 2018). In what follows, we will always use {\tt PyMorph} truncated luminosities.

We will consider E+S0s separately from the full population.  For this, we use the Bayesian Automated morphological classifications (hereafter BAC) of Huertas-Company et al. (2011), because they provide a probability $p$(type) for each object. We can either weight by, or implement hard cuts in, this probability. 

Fixed IMF (Chabrier 2003) $M_*^{\rm SP-Chab}$ estimates for all these galaxies are also available.  These combine the dusty and dust-free $M_*^{\rm SP-Chab}/L$ estimates of Mendel et al. (2014), obtained from integrated multiband photometry, with the truncated S{\'e}rsic $L$ of Meert et al. (2015).  We correct these values for gradients using
\begin{equation}
 M_*^{\rm SP} = M_*^{\rm SP-Chab}\,\frac{1 + g(n,\alpha,\beta)}{1 + g(n,0.39,1)},
 \label{MspSDSS}
\end{equation}
where the values of $\alpha,\beta$ are taken from Table~\ref{tab:ab} as we discuss below. Note that the values of $g$ here are for a S{\'e}rsic profile truncated at $7.5R_e$.

The SDSS pipeline also provides estimates of the velocity dispersion estimated within a circular fiber of radius $\theta_{\rm fiber}=1.5$~arcsec.  For a galaxy at redshift $z$ with S{\'e}rsic index $n$ and half light radius $R_e$, the discussion of the previous section suggests defining $M_*^{\rm dyn}$ using equation~(\ref{Mideal}), with $R_{\rm obs}/R_e = d_{\rm A}(z)\theta_{\rm fiber}$.  However, $R_{\rm obs}$ is approximately $R_e/2$ for the E+S0s in our sample.  In view of our discussion of how gradients will tend to increase the need for dark matter (c.f. Figure~\ref{grad2d}), it is likely that $M_*^{\rm dyn}$ estimated using $\sigma^2_{\rm obs}$ will be biased by dark matter.  For this reason, we first aperture correct $\sigma_{\rm obs}$ to $R_e/8$ using equations~(6) and~(7) of Bernardi et al. (2018): $\sigma(<R)/\sigma(< R_e) = (R/R_e)^{-\gamma(n)}$. (This accounts for a weak dependence of the slope $\gamma$ on the S{\'e}rsic index $n$ instead of being fixed to a constant, $0.06$, as in equation~\ref{sigmacorr}). We then use this corrected value to estimate   
\begin{equation}
  M_*^{\rm dyn} = k(R_e/8,n,\alpha,\beta)\,\left(\frac{R_{\rm obs}}{R_e/8}\right)^{\gamma(n)} \frac{R_e\,\sigma^2_{\rm obs}}{G},
 \label{MdynSDSS}
\end{equation}
where $\alpha$ and $\beta$ are the same values we use in equation~(\ref{MspSDSS}), and the entire analysis uses the truncated S{\'e}rsic profile.  In this respect, our $(\alpha,\beta)=(0,0)$ analysis differs from that in Bernardi et al. (2018) who used a truncated S{\'e}rsic for the light, but whose procedure for estimating $M_*^{\rm dyn}$ was more complex than ours (the net difference in $M_*^{\rm dyn}$ is small).

The aperture correction (i.e., $\gamma$) as well as the  gradient correction (i.e., $k$) depend on galaxy type (i.e., on $n$).  Whereas the aperture correction is a small effect, the gradient correction matters very much.  Further work is necessary to quantify these in data, so here we will explore two simple modifications to the assumption that $\alpha$ and $\beta$ are the same for all galaxies.

\subsection{Type-dependent gradients}
Our first model is motivated by Parikh et al. (2018), who suggest that gradients are smaller in lower mass galaxies.  Therefore, for fixed-IMF $M_*^{\rm SP}$ values that are between 10.3 and 11.2~dex we multiply both $\alpha$ and $\beta$ by
\begin{equation}
  C_* = [\log_{10}(M_*^{\rm SP-Chab}/M_\odot) - 10.3]/(11.2-10.3),
   \label{fmass}
\end{equation}
with $\alpha$ and $\beta$ equal to zero at smaller masses.  This changes the slope of $\Upsilon_*$, but keeps the transition scale beyond which $\Upsilon_*(R)$ becomes constant equal to $\alpha/\beta\sim 0.4 R_e$ for all galaxies. We used the two mass scales identified by Bernardi et al. (2011) as being special:  Various scaling relations change slope at these scales.  We also study a model in which the gradient is weaker at small $\sigma$:  For $\sigma_{e/8}$ between 100 and 250~km~s$^{-1}$ we multiply both $\alpha$ and $\beta$ by
\begin{equation}
  C_\sigma = (\sigma_{e/8}/{\rm kms}^{-1} - 100)/(250-100),
   \label{fsig}
\end{equation}  
with $\alpha$ and $\beta$ equal to zero at smaller $\sigma$.  The threshold values of $\sigma_{e/8}$ are those associated with our threshold values of $M_*^{\rm SP-Chab}$ based on the $\sigma_{e/8}-M_*^{\rm SP-Chab}$ relation.
In what follows, we compare results using $\alpha$ and $\beta$ from Table~\ref{tab:ab}, with and without the mass- or $\sigma$-dependent rescalings we have just described.

\begin{figure*}
 \centering
 \includegraphics[width=0.47\hsize]{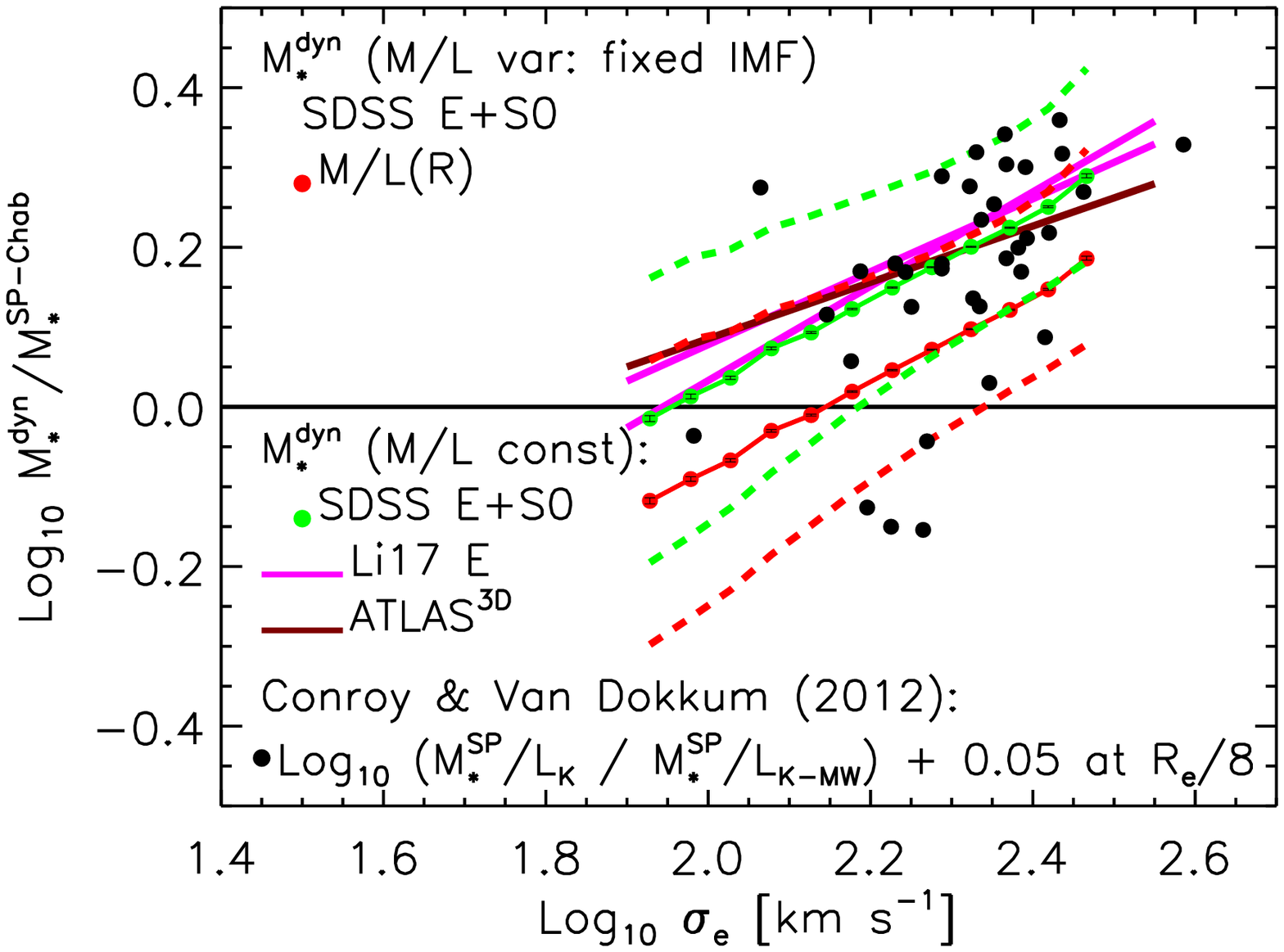}
 \includegraphics[width=0.47\hsize]{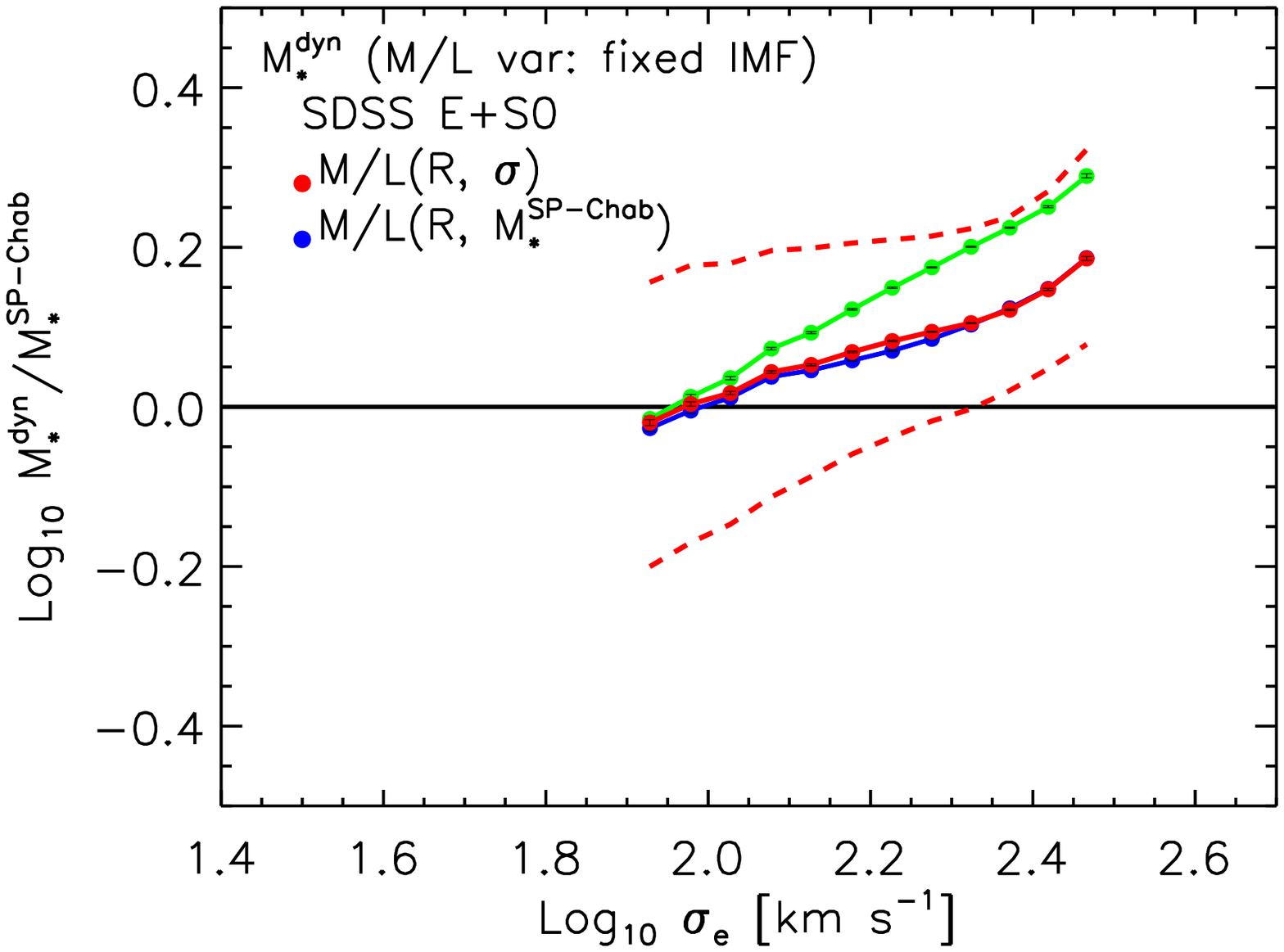}
 \includegraphics[width=0.47\hsize]{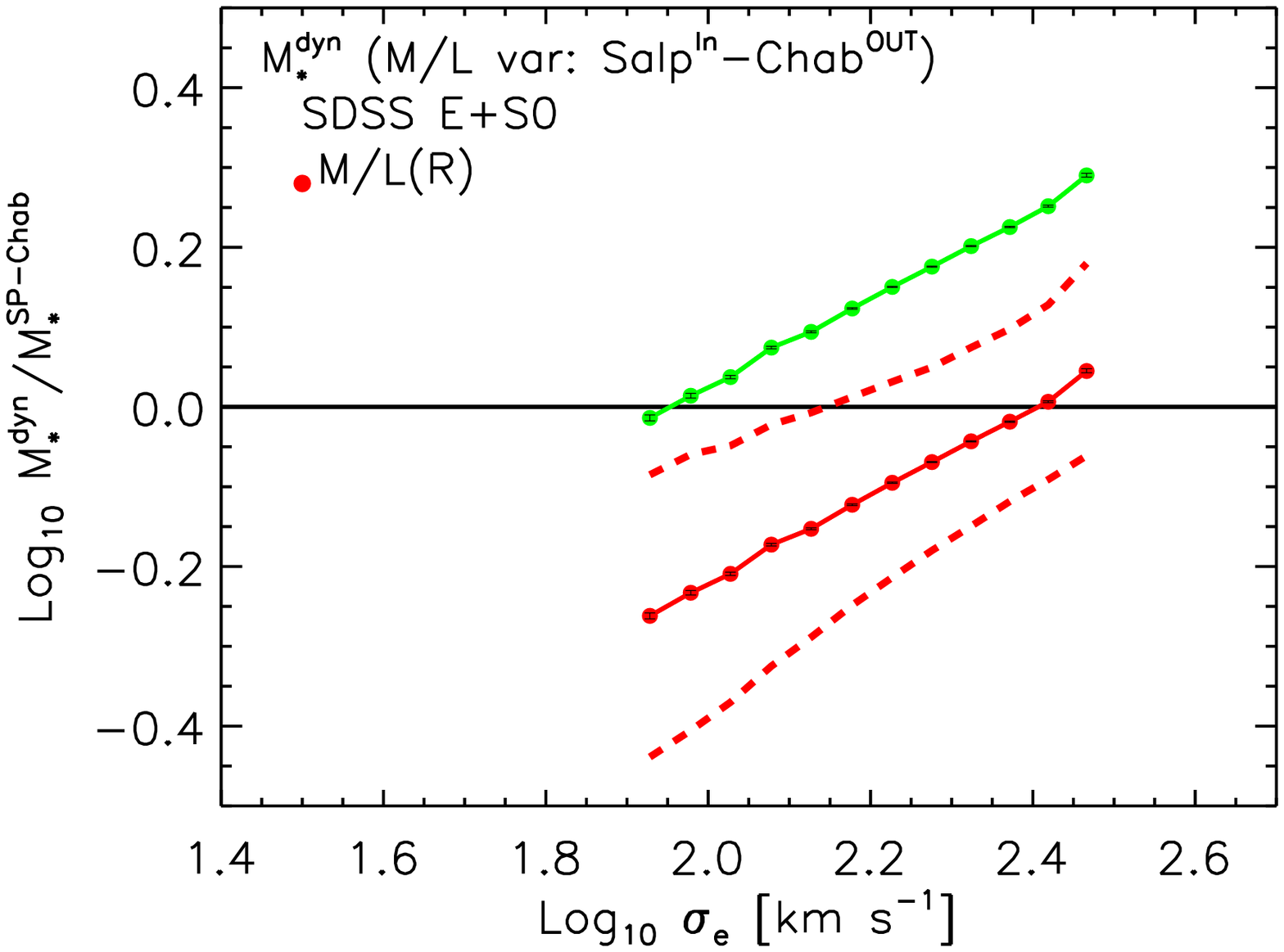}
 \includegraphics[width=0.47\hsize]{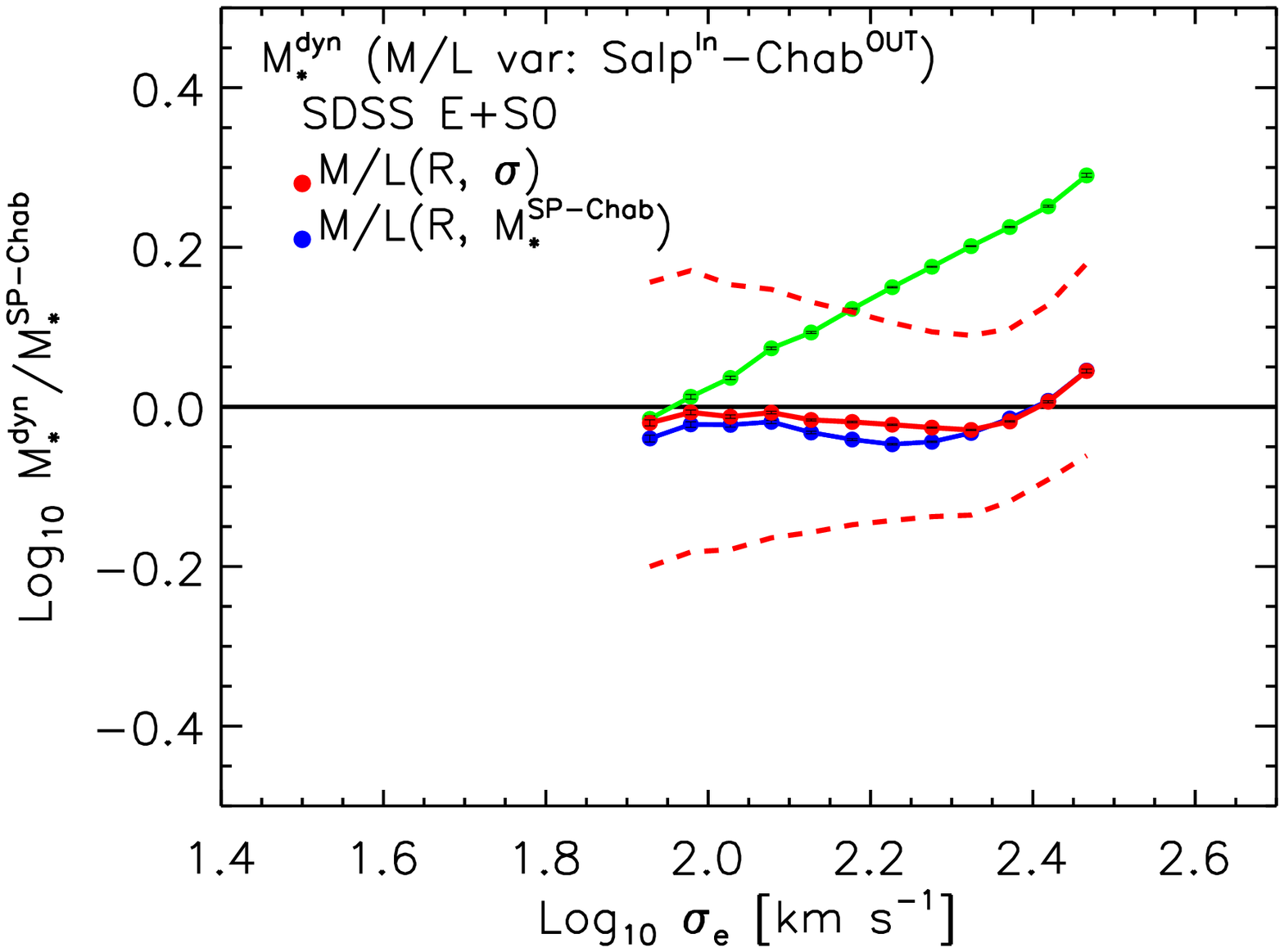}
 \includegraphics[width=0.47\hsize]{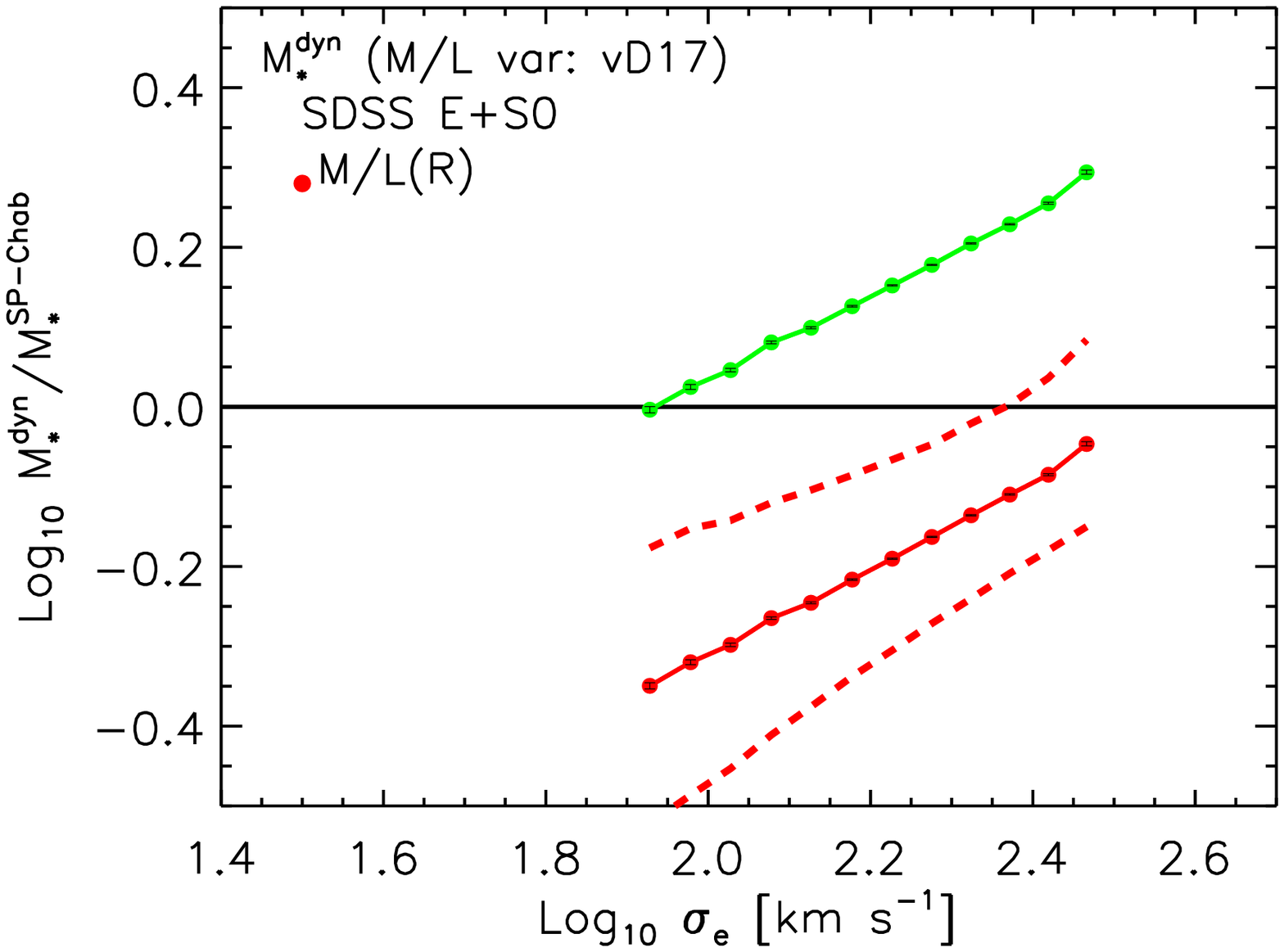}
 \includegraphics[width=0.47\hsize]{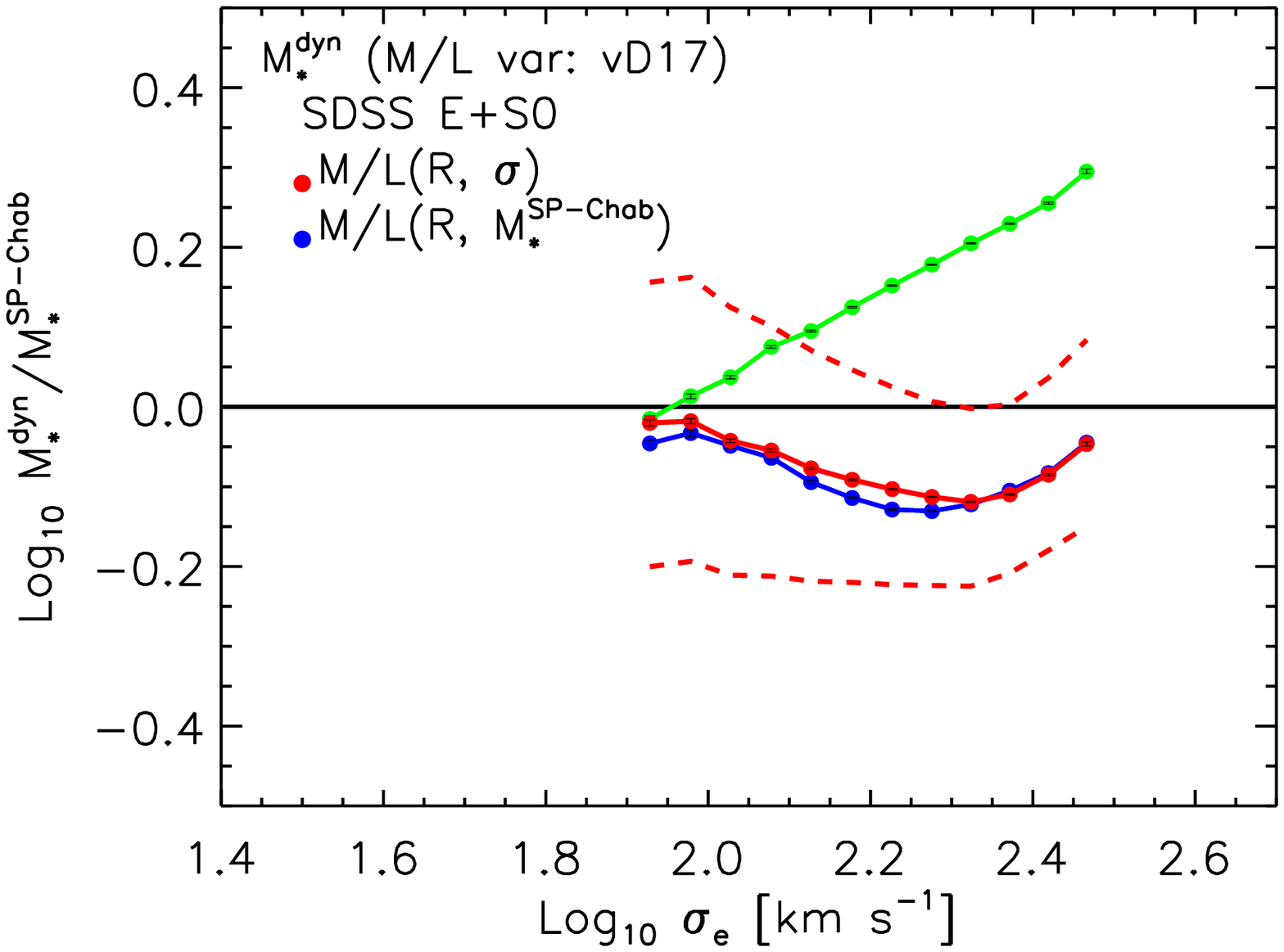}
 \caption{Effect of $\Upsilon_*$ gradients on the $M_*^{\rm dyn}/M_*^{\rm SP}-\sigma$ correlation. {\it Previous work, ignoring $\Upsilon_*$ gradients} (top left panel): Straight lines, same in each panel, show this correlation in the MaNGA (magenta lines; Li et al. 2017, from STARLIGHT and pPXF models) and ATLAS$^{\rm 3D}$ (brown line; Cappellari et al. 2013) datasets if gradients are ignored, and $M_*^{\rm SP}$ estimates assume a Chabrier IMF for all galaxies. Black symbols by Conroy \& van Dokkum (2012; see text for details) show the ratio between stellar mass estimates when the IMF is free to when it is fixed. (We have shifted their $M_*$ estimate to account for the fact their Milky Way IMF is based on Kroupa 2001, which differs by 0.05~dex from Chabrier 2003.) Note, that these estimates were computed in the central regions (at $R_e/8$). Small green dots connected  by a solid line, same in each panel, show the median  $M_*^{\rm dyn}/M_*-\sigma$ relation in SDSS E+S0s if gradients are ignored; dashed lines show the region which encloses 68\% of the objects at each $\sigma_e$ (similar to Bernardi et al. 2018). {\it This work, accounting for $\Upsilon_*$ gradients} (all panels): Small red dots and associated curves result from accounting for gradients using the three models described in Table~\ref{tab:ab}.  Panels on the left assume the gradient strength is the same for all objects, and is due to a fixed (Chabrier) IMF (top -- fixed IMF); to the IMF varying between Salpeter in the center and Chabrier beyond $0.4R_e$ (middle -- Salp$^{\rm IN}$-Chab$^{\rm OUT}$); and to even larger gradients because the center is even more extreme than Salpeter (bottom -- vD17).  Stronger gradients result in greater offsets from the no-gradient case.  Panels on the right show the same models, except that the gradient strength is assumed to decrease at lower masses (equation~\ref{fmass} -- small blue dots) and lower velocity dispersions (equation~\ref{fsig} -- small red dots).  Both show flatter correlations; the slope in the panel on the bottom-most panel is even reversed.  }
 \label{MdMs}
\end{figure*}

Scaling with $\sigma$ rather than $M_*$ is motivated by the fact that IMF-indicators appear to be strongly correlated with $\sigma$ (Conroy \& van Dokkum 2012; Lyubenova et al. 2016).  Moreover, a number of groups have reported that, when gradients are ignored, then $M_*^{\rm dyn}/M_*^{\rm SP-Chab}$ is large if $\sigma$ is large (Cappellari et al. 2013; Li et al. 2017).  This has fueled the argument that $M_*^{\rm dyn}$ estimates are robust, but fixed-IMF $M_*^{\rm SP}$ estimates are biased low if they have not accounted for ($\sigma$-dependent) IMF variations across the population.  However, while van Dokkum et al. (2017) find a large scatter in $\Upsilon_*$ values within $R_e/8$, the range of values at $R_e$ and beyond is much smaller.  If there is a floor to the $\Upsilon_*$ value which is the same for all galaxies, then the ones with small $\sigma$, which Conroy \& van Dokkum (2012) report have small $\Upsilon_*$ even at $R_e/8$, must have small gradients.

If we make a $\sigma$-dependent change to gradients, then we expect that accounting for them will impact the $M_*^{\rm dyn}/M_*^{\rm SP-Chab} - \sigma$ relation.  How this will differ from scaling gradient strength with $M_*$ is less obvious, which is why we believe it is interesting to compare both.

\subsection{Correlation between stellar population and dynamical masses}
Figure~\ref{MdMs} quantifies the impact of $\Upsilon_*$ gradients on the $M_*^{\rm dyn}/M_*^{\rm SP-Chab}-\sigma$ relation using the three pairs of $\alpha,\beta$ given in Table~\ref{tab:ab}.  Assuming the same gradients for all objects changes the amplitude but not the slope of the relation (small red dots in the left hand panels) compared to when one estimates $M_*^{\rm dyn}$ assuming a constant $\Upsilon_*$ (small green dots, same in all panels); reducing the gradient strength at small masses flattens and even reverses the correlation (small blue dots in the right hand panels).  Decreasing the gradients for objects with small $\sigma$ produces very similar results (small red dots in the right hand panels).  

The bottom panels show that the change in amplitude is substantial if we use the largest values of $(\alpha,\beta)$ in Table~\ref{tab:ab} (i.e., Model $=$ vD17); this yields $M_*^{\rm dyn}$ values that are too small compared to the corresponding $M_*^{\rm SP-Chab}$ values.  Top panels show that even the much weaker gradients (Model $=$ fixed IMF) produce noticable effects.  
In all cases, the panels on the right -- in which gradients are weaker at low mass or velocity dispersion -- show shallower or even reversed correlations.  Note that the Figure only shows the effects of changing $M_*^{\rm dyn}$; including the changes to $M_*^{\rm SP}$ as well yields qualitatively similar results, although the quantitative change to $M_*^{\rm SP}$ depends on whether the value of $\Upsilon_*$ which multiplies $L$ was estimated only within the central regions or not.  We illustrate the effect of changes to $M_*^{\rm SP}$ in Section~\ref{MsF}.  

This weakening of the correlation may account for conflicting claims in the literature about the veracity of the IMF-$\sigma$ correlation and its relation to $M_*^{\rm dyn}$.  Conroy \& van Dokkum (2012) report a relation which is steeper than most $M_*^{\rm dyn}/M_*^{\rm SP}$-$\sigma$ relations (e.g. Cappellari et al. 2013; Li et al. 2017).  However, their analysis is really one of the stellar population estimate, based on spectra taken at $R_e/8$; since this is smaller than $0.4R_e$, their results are potentially strongly impacted by gradients.  If the IMF-gradients we have explored are realistic, then our results suggest that the $M_*^{\rm dyn}$ estimates of other groups should be revised downwards, rather than that the $M_*^{\rm SP}$ should be revised upwards.  Of course, actual measurements of how the gradient correlates with $\sigma$ (or $M_*^{\rm SP}$) are needed to really settle the issue, so we hope that our results motivate further work quantifying such trends.

In the meantime, our results illustrate that requiring $M_*^{\rm dyn}=M_*^{\rm SP}$ may provide a useful constraint on gradient strength if there is a lower limit to $\Upsilon_{*0}$ of equation~(\ref{Ur}).  If there is, and one uses it to estimate $M_*^{\rm SP-min}$ from the light, then adding the mass associated with the $\Upsilon_*$ gradient will increase this minimal SP mass estimate by $M_{\rm grad}$ of equation~(\ref{Madded}).  On the other hand, if this additional mass steepens the mass profile compared to the light, it will decrease the dynamical mass estimate (Figure~\ref{grad2d} and related discussion).  As a result, if gradients are too strong, then it is possible that $M_*^{\rm dyn}<M_*^{\rm SP-min}$, which is unacceptable.  The bottom left panel of Figure~\ref{MdMs} shows a model for the $\Upsilon_*$-gradient which results in $M_*^{\rm dyn}<M_*^{\rm SP-Chab}$ (by 0.2~dex on average).  If $M_*^{\rm SP-Chab}\approx M_*^{\rm SP-min}$, then these gradients are unrealistic.

\begin{figure}
 \centering
 \includegraphics[width=0.85\hsize]{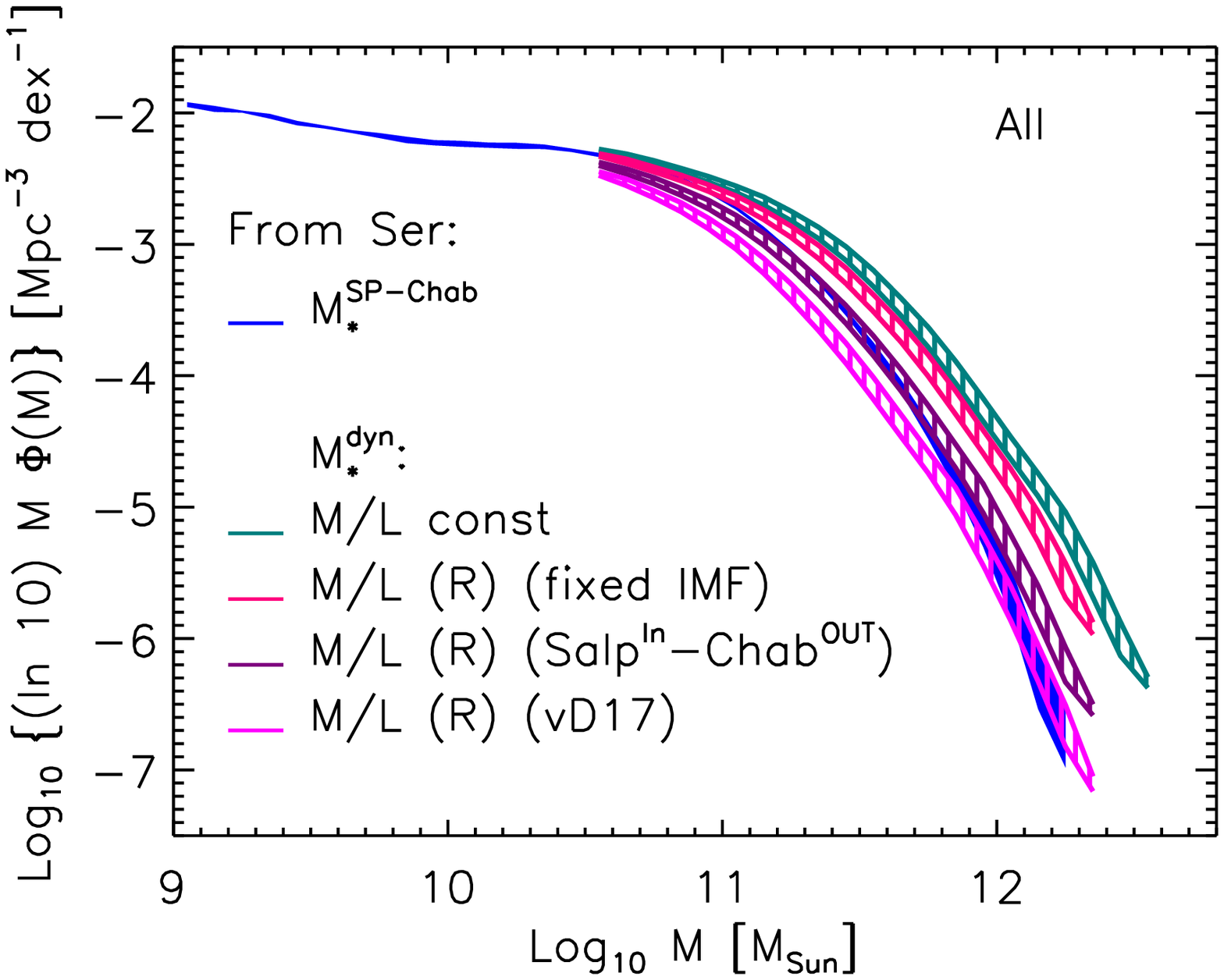}
 \includegraphics[width=0.85\hsize]{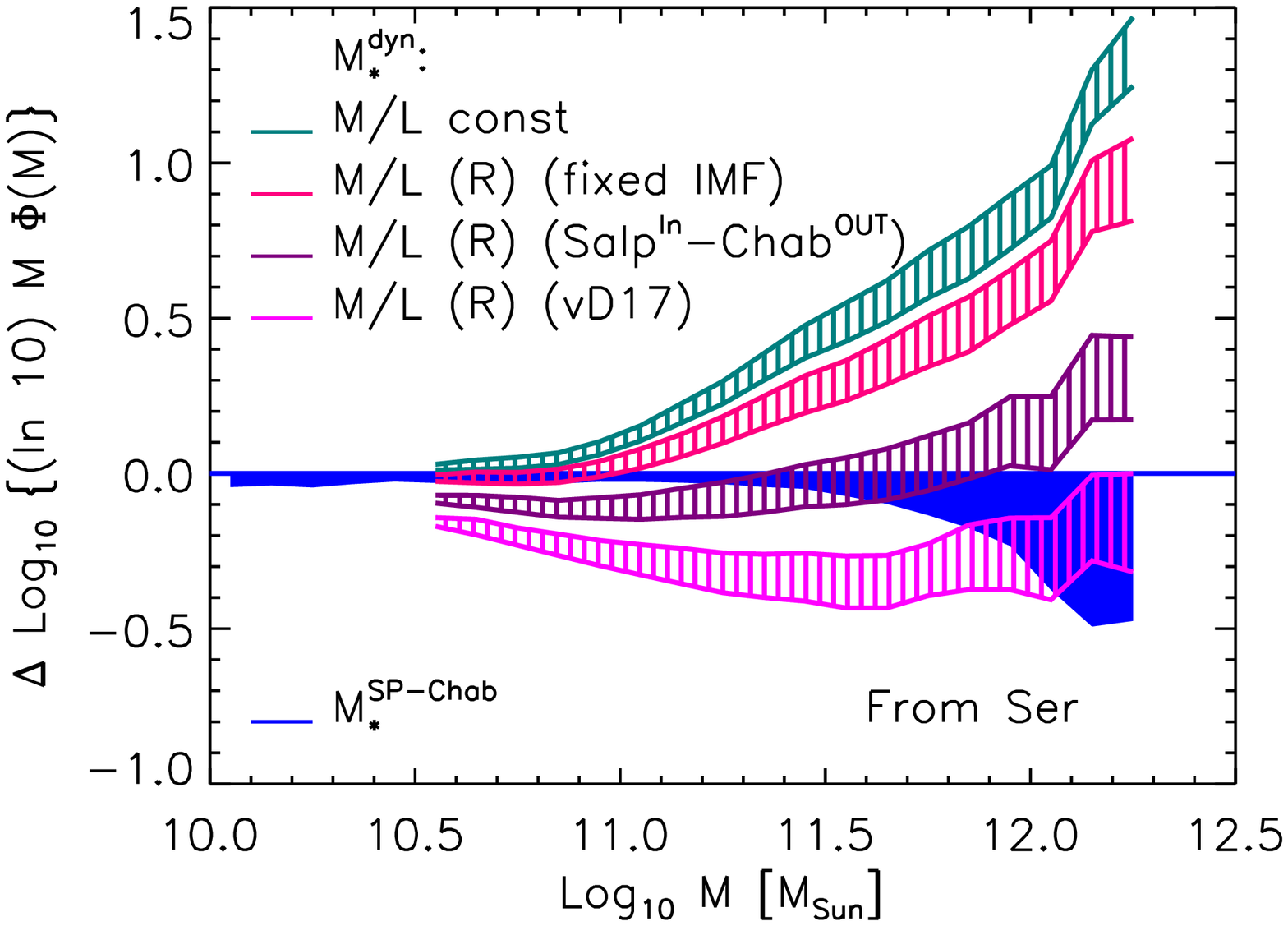}
 \caption{Stellar mass functions for the models shown in the left hand panels of Figure~\ref{MdMs}.  Solid blue region shows the fixed-IMF estimate $\phi(M_*^{\rm SP-Chab})$ and hashed cyan region shows the corresponding $\phi(M_*^{\rm dyn})$ if gradients are ignored (from Bernardi et al. 2018).  Hashed pink, purple and magenta regions (top to bottom) show the result of accounting for $\Upsilon_*$ gradients when estimating $M_*^{\rm dyn}$, as labeled.  Bottom panel shows the same results normalized by a fiducial curve to reduce the dynamic range. Clearly, accounting for $\Upsilon_*$ gradients brings $\phi(M_*^{\rm dyn})$ to better agreement with $\phi(M_*^{\rm SP-Chab})$ -- this is because $M_*^{\rm dyn}$ has been reduced, rather than because $M_*^{\rm SP}$ has been increased. }
   \label{phiMs}
\end{figure}

\begin{figure}
 \centering
 \includegraphics[width=0.85\hsize]{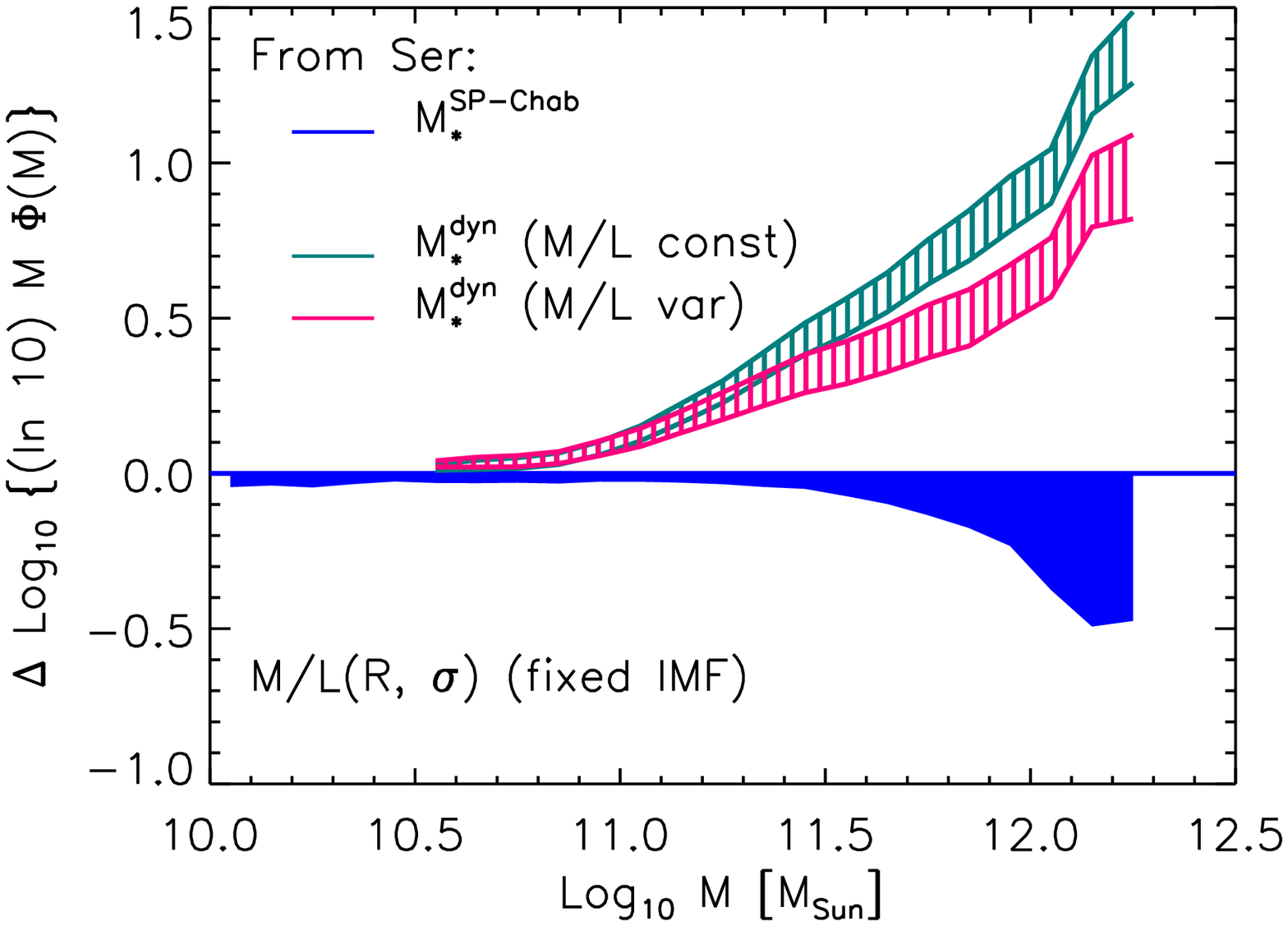}
 \includegraphics[width=0.85\hsize]{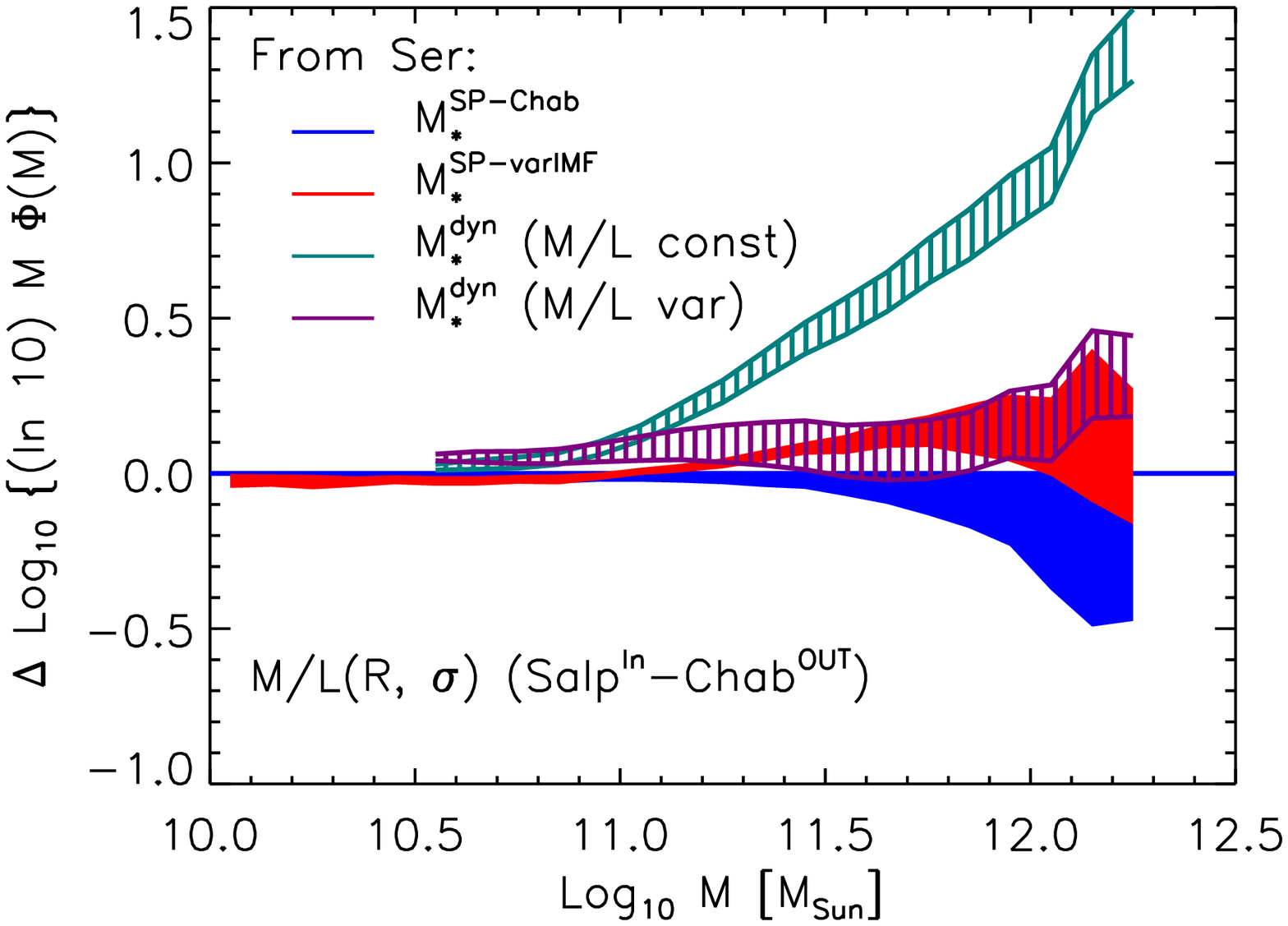}
 \includegraphics[width=0.85\hsize]{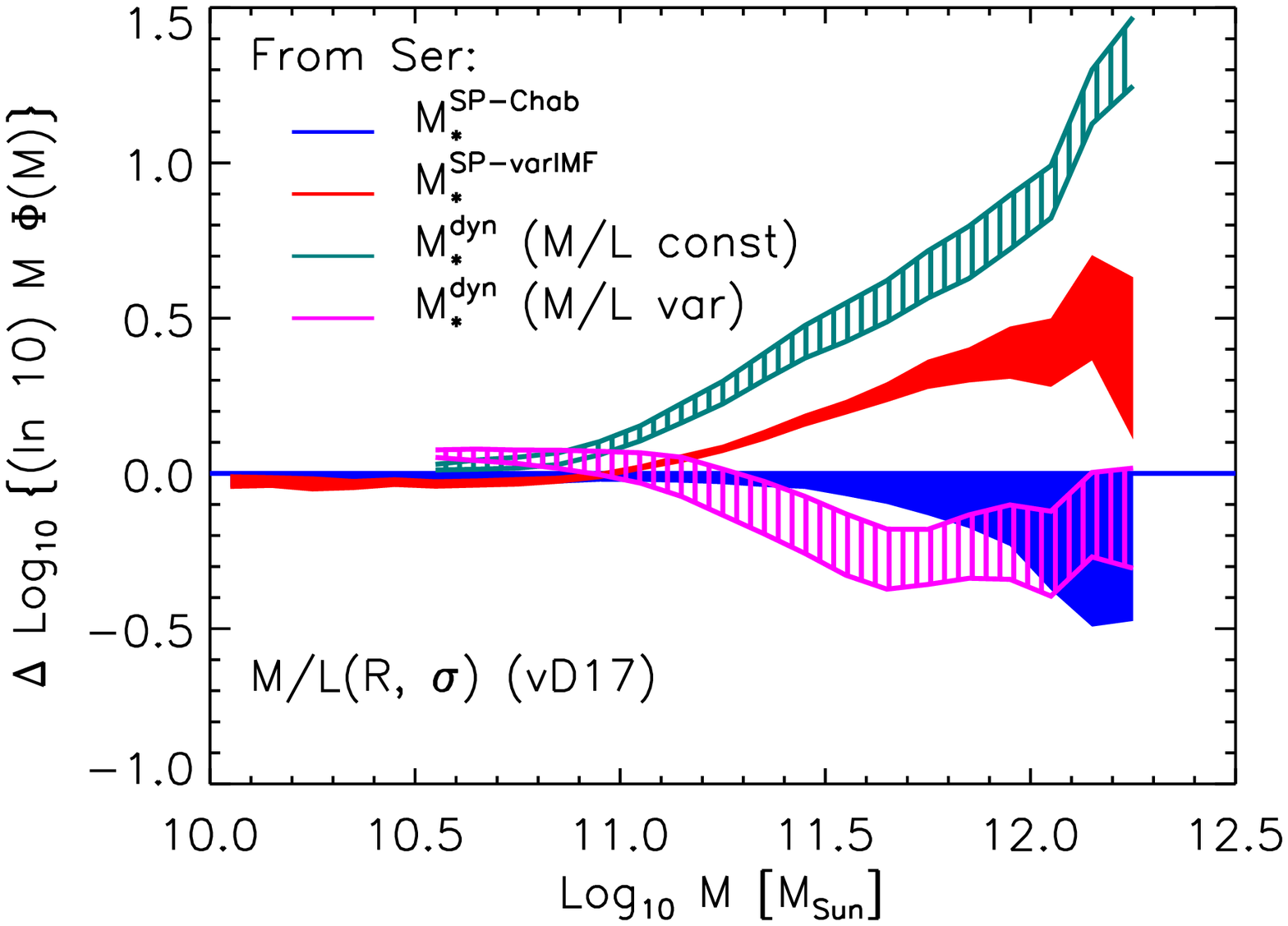}
 \caption{Stellar mass functions for the models shown in the right hand panels of Figure~\ref{MdMs}, divided by a fiducial curve to reduce dynamic range.  Blue solid and hashed cyan regions, same in all panels, show fixed IMF $\phi(M_*^{\rm SP-Chab})$, and $\phi(M_*^{\rm dyn})$ if gradients are ignored.  Red regions show $\phi(M_*^{\rm SP})$ if $\sigma$-dependent gradients have been accounted for.  Pink, purple and magenta show the corresponding $\phi(M_*^{\rm dyn})$ estimates.  Purple and red in middle panel agree because $M_*^{\rm dyn}$ has been reduced, rather than because $M_*^{\rm SP}$ has been increased.  They disagree significantly in bottom panel, where gradients are strongest (also, $\phi(M_*^{\rm dyn}<\phi(M_*^{\rm SP})$, which is opposite compared to top panel).}
   \label{phiMsNorm}
\end{figure}

\subsection{The stellar mass function}\label{MsF}
Including IMF variations across the galaxy population, but ignoring gradients within each galaxy, results in a substantial increase in the inferred stellar mass density that is locked-up in massive galaxies compared to if one assumed a Chabrier IMF for the full population (Bernardi et al. 2018).  In effect, the IMF variation increases the stellar population estimates so that they come into good agreement with the dynamical estimates.  However, the previous subsection showed that accounting for $\Upsilon_*$ gradients within each galaxy will reduce $M_*^{\rm dyn}$; whether $M_*^{\rm SP}$ is also reduced, or must be increased slightly, depends on how it was estimated.  Since these reduction factors can be different, it is not obvious that the two can remain in agreement.  Figure~\ref{MdMs} suggests that if gradients are too strong, then $M_*^{\rm dyn}<M_*^{\rm SP}$ which is obviously problematic (bottom panels).  But for the intermediate strength gradients from our Table~\ref{tab:ab} (Model $=$ Salp$^{\rm IN}$-Chab$^{\rm OUT}$), the two mass estimates can be in reasonable agreement, if gradients are weaker in lower mass galaxies (right middle panel).  

Figures~\ref{phiMs} and~\ref{phiMsNorm} show how these effects manifest in the mass function.  They correspond to the models shown in the left and right hand panels of Figure~\ref{MdMs}:  i.e. same gradient accross the population, or $\sigma$-dependent gradients.  In all panels, solid blue and hashed cyan regions show $\phi(M_*^{\rm SP-Chab})$ and $\phi(M_*^{\rm dyn})$ if gradients are ignored (from Bernardi et al. 2018); except for the top panel of Figure~\ref{phiMs} we have normalized the results by a fiducial curve to more clearly illustrate the differences.  The large differences between the blue and cyan regions, especially at large masses, are usually attributed to problems with $M_*^{\rm SP-Chab}$ rather than $M_*^{\rm dyn}$, but our analysis suggests that accounting for gradients might lead to the opposite conclusion.

The pink, purple and magenta regions (top to bottom) in Figure~\ref{phiMs} show the effect of $\Upsilon_*$ gradients when they are the same across the population.  If there are no IMF gradients, so the only $\Upsilon_*$ gradients are those associated with a fixed IMF, then $M_*^{\rm SP}=M_*^{\rm SP-Chab}$ (see equation~\ref{MspSDSS}), and $\Upsilon_*$ gradients only modify $M_*^{\rm dyn}$.  Since they are weak, the decrease in $M_*^{\rm dyn}$ is small:  this brings the hashed cyan region down to the hashed pink region.  This is why such gradients are usually ignored.  However, the stronger IMF-driven gradients produce more dramatic changes to $\phi(M_*^{\rm dyn})$.  

The three panels in Figure~\ref{phiMsNorm} show how gradients impact the stellar mass function for the three models shown on the right hand side of Figure~\ref{MdMs}:  these use the three sets of $\alpha,\beta$ given in Table~\ref{tab:ab}, but scaled so that objects with small $\sigma$ have smaller gradients (recall that scaling with mass gives similar results).  The stellar population and dynamical mass results are similar only in the middle panel.  When gradients are too weak (top panel), then $M_*^{\rm SP}<M_*^{\rm dyn}$, whereas the opposite is true when gradients are too strong (bottom panel).  In the middle panel however, the two estimates are slightly larger than the fixed-IMF estimate $\phi(M_*^{\rm SP-Chab})$ which is currently in the literature (Bernardi et al. 2017a, 2018), but substantially smaller than the dynamical estimate associated with ignoring gradients.  I.e., agreement between the two stellar mass functions is achieved by decreasing $M_*^{\rm dyn}$ so it matches $M_*^{\rm SP-Chab}$ rather than the other way round.  

Clearly, an accurate census of the stellar mass density requires that gradients in $\Upsilon_*$ be well-quantified.

\section{Discussion}
Observations suggest higher stellar mass-to-light ratios $\Upsilon_*$ in the central regions of galaxies.  We illustrated a number of ways in which such gradients in $\Upsilon_*$ impact stellar population and dynamical mass estimates of galaxies.

Ignoring such gradients leads to overestimates of the dynamical stellar mass (Figure~\ref{grad2d} and related discussion, as well as Figure~\ref{grad3d} and equation~\ref{MLtheory}), with $M_*^{\rm dyn}$ being more strongly biased (Figures~\ref{gradPS} and~\ref{biasMdyn}).  Since gradients are not well-quantified in the literature, we illustrated their effects using three models (equation~\ref{Ur} with Table~\ref{tab:ab}). We noted that, if there is a minimum value to $\Upsilon_*$, then requiring agreement between $M_*^{\rm dyn}$ and $M_*^{\rm SP}$ limits the allowed gradient strength (bottom left panel of Figure~\ref{MdMs} and related discussion).  We also explored the possiblitity that $\Upsilon_*$-gradient strength depends on galaxy type -- e.g. if gradients are larger at higher velocity dispersion or stellar mass.  Gradients can modify the $M_*^{\rm dyn}/M_*^{\rm SP}-\sigma$ substantially (Figure~\ref{MdMs}).

Of the three simple models for gradients which we explored, only the one where gradients are driven by a transition from Salpeter in the inner regions to Chabrier beyond $0.4R_e$ produces agreement between $M_{\rm dyn}$ and $M_*^{\rm SP}$, and then, only if the gradient strength is weaker in less massive galaxies. In this model, IMF-driven gradients bring $M_*^{\rm dyn}$ and $M_*^{\rm SP}$ into agreement, not by shifting $M_*^{\rm SP}$ upwards by invoking constant bottom-heavy IMFs, as advocated by a number of recent studies (e.g. Li et al. 2017), but by revising $M_*^{\rm dyn}$ estimates in the literature (e.g. Cappellari et al. 2013) downwards.

The reduction in $M_*^{\rm dyn}$ has the effect of shifting the cutoff in $\phi(M_*^{\rm dyn})$ to masses that are approximately $2\times$ smaller (Figures~\ref{phiMs} and~\ref{phiMsNorm}).  In our Salp$^{\rm IN}$-Chab$^{\rm OUT}$ model, agreement between $\phi(M_*^{\rm SP})$ and $\phi(M_*^{\rm dyn})$ is achieved by decreasing $M_*^{\rm dyn}$ so it matches $M_*^{\rm SP-Chab}$ rather than the other way round.  As a result, the shape of the mass function is close to $\phi(M_*^{\rm SP-Chab})$ which is provided in tabular form by Bernardi et al. (2018).  Therefore, an accurate census of the stellar mass density requires gradients in $\Upsilon_*$ to be well-quantified.

Smaller dynamical stellar masses (compared to previous $M_*^{\rm dyn}$ estimates, such as those provided by the JAM-analyses of Cappellari et al. 2013 and Li et al. 2017 for the ATLAS$^{\rm 3D}$ and MaNGA samples, respectively) also mean that dark matter must play a more dominant role than previously thought (Figure~\ref{grad2d}). Although the details will also depend on velocity anisotropy, our analysis of the simplest case suggest these will be sub-dominant (Figures~\ref{gradPS} and~\ref{biasMdyn}). In turn, this impacts discussions of whether or not dark matter halos must have contracted adiabatically because of the baryons which cooled and formed stars in their centers (Newman et al. 2015; Shankar et al. 2017, 2018).

More centrally concentrated baryonic mass also impacts inferences about the black hole sphere of inference, especially if the S{\'e}rsic index $n$ is small. Without gradients, $\sigma_p$ is predicted to decrease at small $r$, so if an increase is observed, it may be attributed to a central black hole.  Anisotropic velocity dispersions complicate the discussion (Binney \& Mamon 1982; van der Marel 1994), but since $\Upsilon_*$-gradients can also change the small scale slope, they are an additional complication.  However, if black hole mass fractions are as small as Shankar et al. (2016) suggest, then they only matter on much smaller scales than those considered here (equation~\ref{mbh} and related discussion).  

Finally, more centrally concentrated baryons will change estimates of the radial acceleration relation which is sometimes used to constrain modifications to standard gravity (Chae et al. 2014; Tortora et al. 2014; Chae \& Gong 2015; Lelli et al. 2016; Posacki et al. 2016; Tortora et al. 2018; Chae et al. 2018; Appendix~A2 here).  For all these reasons, we hope that our work stimulates efforts to quantify how gradients in the stellar mass-to-light ratio depend on galaxy type.  

Our final point is a philosophical one.  If $\Upsilon_*$ gradients can be ignored, then $M_*^{\rm dyn}$ provides an estimate of the stellar mass that is independent of stellar population modeling.  This independence is sometimes used to argue that the extra work required to estimate $M_*^{\rm dyn}$ is well-worth the trouble, since $M_*^{\rm dyn}$ is insensitive to the details of the stellar population which can lead to systematic biases in $M_*^{\rm SP}$.  However, at the present time, $\Upsilon_*$ gradients are typically estimated from stellar population modeling, so the need to include gradients when estimating $M_*^{\rm dyn}$ removes this independence.  As independent checks on mass-estimates are desirable, our work provides strong motivation for alternative probes of the IMF, such as the microlensing method of Schechter et al. (2014).

\subsection*{Acknowledgements}
We are grateful to the referee for an encouraging report, and to the USTA for its generosity when this work was completed.

\appendix
\section{Simple model for effects of $M_*/L$ gradients and a Hernquist profile}
We use Hernquist's (1990) profile to illustrate the effect of gradients in the stellar mass-to-light profile on the estimated stellar population and dynamical mass.  

\subsection{Newtonian gravity}
Suppose that the three dimensional distribution of the light follows a Hernquist (1990) profile
\begin{equation}
  \rho_{\rm L}(r) = \frac{L/R_H^3}{2\pi\,(r/R_H)(1 + r/R_H)^3} \equiv
  \frac{L/R_H^3}{2\pi\,r_H(1 + r_H)^3},
  \label{rhoH}
\end{equation}
where $L$ is the total luminosity, $R_H$ is a characteristic radius and $r_H \equiv r/R_H$.  The luminosity within $r$ is 
\begin{equation}
 L(<r) = L\,\frac{r_H^2}{(1 + r_H)^2}.
\end{equation}
If $M_*/L$ increases towards the center as $\Upsilon_{*0}\,(1 + f/r_H)$, so the mass density profile is
\begin{equation}
  \rho_{\rm M}(r) = \Upsilon_{*0}\,\frac{f+r_H}{r_H}\,\rho_{\rm L}(r),
\end{equation}
then
\begin{equation}
 M(<r) 
  = \Upsilon_{*0}\,L(<r)\,(1 + f + 2f/r_H).
 \label{MLtheory}
\end{equation}
Thus $M_*(<r)/L(<r) = \Upsilon_{*0}\,(1 + 3f)$ at $r=R_H$
and $\Upsilon_{*0} \,(1 + f)$ at $r\gg R_H$.
Therefore, the total mass is $1+f$ times larger than $\Upsilon L$, the value if there were no gradient.

There are two natural proxies for the mass estimated by stellar population models when gradients are ignored.  One is the light-weighted mass-to-light ratio:  $M_\infty = \Upsilon_{*0} L (1+f)$.  However, if $M_*^{\rm SP}/L$ was estimated from the light within the projected half light radius $R_e\approx 1.8153 R_{\rm H}$, then it might be more natural to use $2\times$ the mass projected within $R_e$, which is $\Upsilon_{*0} L (1 + 1.816f)$.  This will generically overestimate $M_\infty$.  

We turn now to dynamical mass estimates.  For these, if we assume the velocity dispersion is isotropic, then the Jeans equation implies that 
\begin{equation}
  \rho_M(r)\,\sigma^2(r)
  = \int_r^\infty \frac{{\rm d}r}{r}\,\rho_M(r)\,\frac{GM(<r)}{r} .
\end{equation}  
The integral can be done analytically, and yields 
\begin{align}
 & \rho_{\rm M}(r)\frac{\sigma^2(r)}{GM_\infty/R_{\rm H}}
  = \int_r^\infty \frac{{\rm d}r}{r}\,\rho_M(r)\,\frac{M(<r)/M_\infty}{r/R_{\rm H}}\nonumber\\
 &= -\frac{\rho_{\rm M}(r)/(1+f)}{12 (f+r_{\rm H}) (r_{\rm H} + 1)}
 \Bigl(f^2 (5 r_{\rm H} (5 r_{\rm H} A + 12) - 12)\nonumber\\
 & \qquad - 12 \bigl(f\, (25 f - 14) + 1\bigr)\, r_{\rm H}^2\, (r_{\rm H} + 1)^4 \,\ln(1/r_{\rm H} + 1) \nonumber\\
 & \qquad - 2 f r_{\rm H} (7 r_{\rm H} A + 18) + r_{\rm H}^2 A\Bigr),
 \label{sigma3d}
\end{align}
where $M_\infty\equiv \Upsilon_{*0} L\,(1+f)$ and 
 $A = 2 r_{\rm H} \bigl(3 r_{\rm H} (2 r_{\rm H} + 7) + 26\bigr) + 25.$
If we add a central black hole of mass $M_\bullet$, then
\begin{align}
 & \frac{M_\bullet}{M_\infty}\,\frac{\rho_{\rm M}(r)}{6 r_{\rm H} (f+r_{\rm H})/(1+r_{\rm H})}\, \nonumber\\
 & \Bigl(2 (3 - 5f) (r_{\rm H} + 1)^3 r_{\rm H}^2 \ln(1 + 1/r_{\rm H}) \nonumber\\
 & \quad + f \bigl(5 r_{\rm H} (2 r_{\rm H} + 1) B + 2\bigr) - 3r_{\rm H} (2 r_{\rm H} + 1) B\Bigr),
 \label{mbh}
\end{align}
where $B = 6 r_{\rm H} (r_{\rm H} + 1) - 1$, must be added to the expression above.

The light-weighted projected dispersion at projected radius $R$ is given by equation~(\ref{sigmap}) in the main text.  Models with a constant mass to light ratio have $\rho_{\rm M}(r) = \Upsilon_{*0} \rho_{\rm L}(r)$, so the shape of the light profile determines the shape of $\sigma_p$.  The dynamical mass is then estimated by finding that $\Upsilon_{*0}$ which gives the observed value of $\sigma_p$:  hence, the value of $\Upsilon_{*0}$ from stellar population modeling is irrelevant.

\begin{figure}
 \centering
 \includegraphics[width=0.9\hsize]{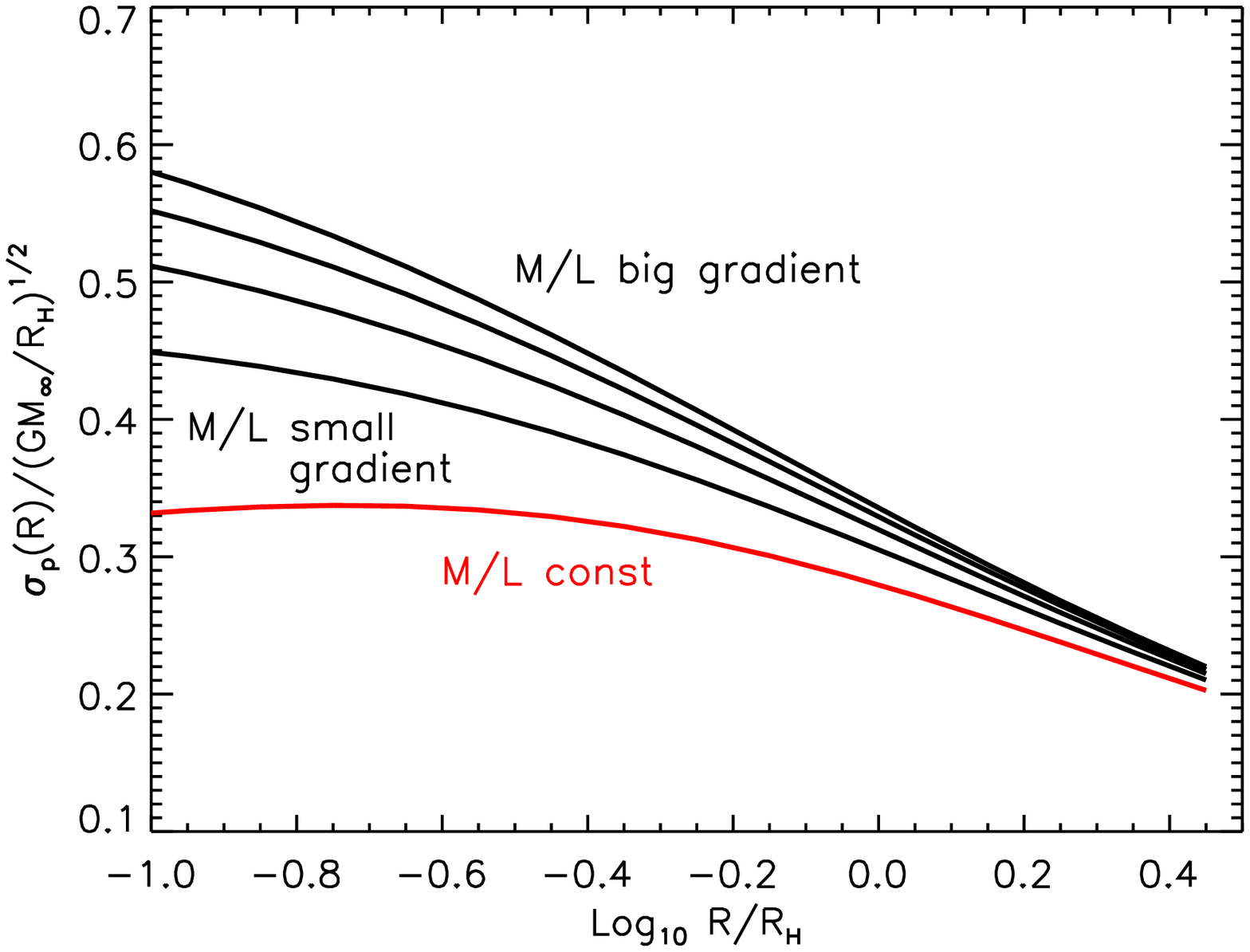}
 \caption{Effect of gradients in the mass-to-light ratio on the light-weighted projected velocity dispersion if the light distribution follows a Hernquist profile with scale radius $R_{\rm H}$ and the mass-to-light ratio increases towards the center as $1 + fR_{\rm H}/r$.  Curves, from bottom to top, show $\sigma_p$ for $f=(0,0.25,0.5,0.75,1)$:  gradients steepen the $\sigma_p$ profile.  When normalized to the dispersion observed on scale $R$, the stellar dynamical mass estimate is proportional to the inverse of the square of the quantity shown on the y-axis.
 }
 \label{grad3d}
\end{figure}

However, when there are gradients, then stellar population modelling is {\em required} to get $\rho_{\rm M}/\rho_{\rm L}$, so the dynamical mass estimate depends on the SP modeling (upto an overall normalization factor).  I.e., the SP modelling is needed to estimate the shape, after which the normalization is given, as before, by fitting to data.  For example, suppose we normalize using the observed $\sigma_p$ on some scale $R_{\rm obs}$.  Then,
\begin{equation}
  M_{\rm dyn} = k(R_{\rm obs},f)\, R_{\rm H}\,\sigma^2_{\rm obs}/G, 
\end{equation}
where $k(R_{\rm obs},f) = (GM_\infty/R_{\rm H})/\sigma^2_p(R_{\rm obs})$ is the inverse of the (square of the) value of the curve with the relevant value of $f$ shown in Figure~\ref{grad3d}.  

If we assume there are no gradients when, in fact, there are, then we will use the red curve to match the observations instead of the correct one.  This will have two consequences.  First, the resulting stellar mass estimate will be too high by a factor of $k(R_{\rm obs},f)/k(R_{\rm obs},0)$.  This can be a significant overestimate if $R_{\rm obs}<R_{\rm H}$.  For example, if $R_{\rm obs} = R_e/8 \approx 0.2R_{\rm H}$ and $f=0.5$ then this factor is $(0.475/0.33)^2\approx 2$.  

The second effect has to do with the fact that the curves with larger $f$ are steeper.  In practice, observed $\sigma(R)$ profiles are shallower than the ones predicted by a Jeans equation analysis of the light (+ constant mass to light ratio).  The difference is attributed to dark matter.  If we were to slide the red curve upward so it matched a black one at some $R=R_{\rm obs}$, then it would lie above the black beyond beyond $R_{\rm obs}$.  This would reduce the amount of dark matter which is required on scales beyond $R_{\rm obs}$.  As a result, dynamical mass estimates which ignore gradients generically bias one towards underestimating the amount of dark matter on scales where the light is seen.  This will lead one to underestimate the amount of adiabatic contraction of the dark matter that the baryons in the central regions caused.

\subsection{Modified Newtonian dynamics}
For MOND, if $a_{\rm N}\equiv GM(<r)/r^2$ then the acceleration becomes $a_{\rm N}(r)\, f(a_{\rm N}/a_0)$.  There are a number of parametrizations of this scaling, but to illustrate our results, we will use $f_{\rm Bek}(x) = 1 + 1/\sqrt{x}$.  For a Hernquist light profile,
\begin{equation}
  a_{\rm N}(r) = \frac{G\,\Upsilon_{*0} L/R_{\rm H}^2}{(1+r/R_{\rm H})^2}
          \equiv \frac{a_{\rm H}}{(1 + r_{\rm H})^2},
\end{equation}
so $f_{\rm Bek} = 1 + (1+r_{\rm H})\sqrt{a_0/a_{\rm H}}$.  Thus, the three dimensional velocity dispersion satisfies 
\begin{equation}
  \rho_{\rm L}(r)\sigma^2(r) = R_{\rm H}a_{\rm H}\,\frac{L/R_{\rm H}^3}{2\pi}\,
    \int_{r_{\rm H}}^\infty \frac{{\rm d}r}{r}\,
    \left[\frac{1}{(1+r)^5} +\frac{\sqrt{a_0/a_H}}{(1+r)^4}\right].
\end{equation}
The term proportional to $a_0/a_{\rm H}$ is a correction to the Newtonian expression given earlier (equation~\ref{sigma3d} with $f=0$); it only matters when $a_{\rm H}\ll a_0$.  The integral for this correction factor can be done analytically, yielding $(G\,\Upsilon_{*0} L/R_{\rm H})\,(L/R_{\rm H}^3)/2\pi$ times 
\begin{equation*}
   \sqrt{\frac{a_0}{a_{\rm H}}}
  \Bigl(\ln(1/r_{\rm H} + 1) - \frac{3 r_{\rm H} (2 r_{\rm H} + 5) + 11}{6\,(1+r_{\rm H})^3}\Bigr).
\end{equation*}
The presence of $a_{\rm H}\propto\Upsilon_*$ under the square root sign means that $\sigma(r)$ is not linearly proportional to $\Upsilon_*$.  Therefore, in contrast to the Newtonian case, as one changes the normalization factor $\Upsilon_*$, the shape of $\sigma(r)$ also changes.  This is because, in the Newtonian case, one treats the discrepancy between what the light predicts and what is observed by add an independent term, $M_{\rm dm}(<r)$ to the total mass.  In MOND, what matters is the ratio $a_0/a_{\rm H}$, so the amount to be `added' depends on what is present.  

With gradients, 
\begin{equation}
  a_{\rm N}(r)
       \equiv a_{\rm H}\,\frac{(1 + f + 2f/r_{\rm H})}{(1 + r_{\rm H})^2},
\end{equation}
so
 $f_{\rm Bek} = 1 + \sqrt{a_0/a_{\rm H}}\,(1+r_{\rm H})/\sqrt{1 + f + 2f/r_{\rm H}}$.  This complicates the integral for $\sigma(r)$, but not the fact that the shape of $\sigma(r)$ depends on $\Upsilon_*$.  

For example, when $f=1$ then $\rho_{\rm M}(r) = \rho_{\rm L}(r)\, (1 + r_{\rm H})/r_{\rm H}$ so $M(<r) = M_\infty\, r_{\rm H}/(1+r_{\rm H})$ with $M_\infty = 2\Upsilon_{*0} L$, and
\begin{align}
  & \frac{\sigma^2(r)}{GM_\infty/R_{\rm H}} = 6\,r_{\rm H}^2\,(1 + r_{\rm H})^2\ln(1 + 1/r_{\rm H}) \nonumber\\
  & \qquad \qquad - \frac{(1 + 2r_{\rm H})\Bigl[6r_{\rm H}(1+r_{\rm H}) - 1\Bigr]}{2} \\
  & \qquad + \sqrt{\frac{a_0}{2a_{\rm H}}}\,\frac{1 - 2r_{\rm H}\,\Bigl[4r_{\rm H}\,(2r_{\rm H} - 2D + 3) - 8D + 3\Bigr]}{3/2D}, \nonumber
\end{align}
where $D = \sqrt{r_{\rm H}(1+r_{\rm H})}$.

\end{document}